\documentclass[%
reprint,
 amsmath,amssymb,
 aps,
pre]{revtex4-2}

\usepackage{graphicx}
\usepackage{dcolumn}
\usepackage{bm}
\usepackage{tikz}
\usetikzlibrary{positioning}
\usepackage{graphicx}
\usepackage{subcaption}
\usepackage{caption}
\usepackage{float}
\usepackage{booktabs} 
\usepackage{siunitx} 
\usepackage{multirow}
\usepackage{adjustbox} 
\usepackage{placeins}
\usepackage{enumitem}
\usetikzlibrary{positioning, calc, arrows.meta, shapes.geometric}

\begin{document}

\preprint{APS/123-QED}

\title{Uncertainty in AI-driven Monte Carlo simulations}

\author{
Dimitrios Tzivrailis\(^{1,2}\), 
Alberto Rosso\(^{2}\),
Eiji Kawasaki\(^{1}\) 
}

\affiliation{\(^1\) Université Paris-Saclay, CEA, List, F-91120 Palaiseau Cedex, France}
\affiliation{\(^2\) LPTMS, CNRS, Université Paris-Saclay, 91405 Orsay, France}

\date{May 26, 2026}

\begin{abstract}

In the study of complex systems, evaluating physical observables often requires sampling representative configurations via Monte Carlo techniques. These methods rely on repeated evaluations of the system’s energy and force fields, which can become computationally expensive. To accelerate these simulations, deep learning models are increasingly employed as surrogate functions to approximate the energy landscape or force fields. However, such models introduce epistemic uncertainty in their predictions, which may propagate through the sampling process and affect the simulation’s macroscopic behavior. In our work, we present the Penalty Ensemble Method (PEM) to quantify epistemic uncertainty and mitigate its impact on Monte Carlo sampling. Our approach introduces an uncertainty-aware modification of the Metropolis acceptance rule, which increases the rejection probability in regions of high uncertainty, thereby enhancing the reliability of the simulation outcomes.

\end{abstract}

\maketitle

\section{Introduction}

The integration of artificial intelligence (AI), particularly machine learning, into scientific methodologies is transforming physics and chemistry. This synergy offers new paradigms for modeling complex phenomena, combining established physical laws with data-driven approaches \cite{carleo_machine_2019, duignan_potential_2024}. However, such powerful data-driven models inherently introduce epistemic uncertainty into their predictions \cite{zhou_survey_2022}.

The investigation of complex systems often necessitates sampling configurations via Monte Carlo techniques, which depend on computationally intensive energy and force field evaluations \cite{paquet_molecular_2015}. To accelerate these simulations and enable studies of larger, intricate systems, deep learning models are increasingly used as efficient surrogate functions \cite{chmiela_towards_2018, bono_performance_2025}. This integration promises to significantly enhance physical modeling by replacing resource-intensive computations with rapid approximations \cite{noe_machine_2020}.

However, this hybridization introduces significant challenges. Machine learning models yield predictions with statistical properties distinct from those of direct physical calculations, containing inherent "noise" due to the finite and often limited observational data used for training. To effectively characterize these AI models as statistical estimators, robust methods for uncertainty quantification are essential, an area that has begun to be explored \cite{grasselli_uncertainty_2025, imbalzano_uncertainty_2021, lin_searching_2021, swinburne_parameter_2025, peterson_addressing_2017, psaros_uncertainty_2023, imbalzano_uncertainty_2021, perez_uncertainty_2025}. Nevertheless, AI models such as Neural Network Interatomic Potentials are known to produce unstable simulations, frequently leading to irreversible entry into unphysical regions of phase space \cite{fu_forces_nodate}. This propagation of predictive noise and the resulting incompatibility of proposed configurations with the simulation, especially as system size increases, represents a major bottleneck for the robust application of deep learning in scientific computing \cite{focassio_performance_2024}. Such perturbations can profoundly affect the macroscopic behavior of complex, stochastic systems.

This study directly addresses the fundamental challenge of epistemic uncertainty arising from the integration of deep learning models into physical simulations. When a computationally expensive physical energy calculation is approximated by a neural network prediction, a residual error is inevitably introduced. This residual can significantly influence simulation results, introducing bias into macroscopic physical quantities and corrupting the simulation even if its mean is nominally zero, leading to inaccurate conclusions about the system
properties and dynamics. Even seemingly benign Gaussian noise in a surrogate model, if unaddressed, can lead a Monte Carlo simulation to sample an incorrect distribution, thereby biasing the derived physical observables \cite{Ceperley_1999}. This work presents two primary contributions. First, we show how such bias is introduced into physical observables when Monte Carlo simulations are driven by neural network potentials, even when the surrogate model exhibits good performance on test data. Second, we propose the Penalty Ensemble Method (PEM) as a robust solution to mitigate this bias. PEM quantifies epistemic uncertainty and directly integrates this information into Monte Carlo sampling through an uncertainty-aware modification of the Metropolis acceptance rule. This modification selectively increases rejection probability in regions of high predictive uncertainty, a novel mechanism that enhances the reliability and physical consistency of simulation outcomes and
prevents the sampling of out-of-distribution states.

In the following, we approximate the action of the two-dimensional $\phi^4$ model using a Residual Convolutional Neural Network (RCNN). This RCNN surrogate is integrated into the Metropolis algorithm, replacing the exact analytical expression for the model's action. The inherent approximation error of the RCNN directly influences the operation of the Metropolis algorithm. We generate samples to compute critical physical observables, exploring both far and near-phase transition regimes. Our findings clearly demonstrate how the RCNN's approximation error biases the sampling process, leading to significant deviations in the estimations of these physical observables compared to those obtained from conventional Metropolis simulations. To counteract this, our work proposes the PEM to estimate the model's uncertainty using an ensemble technique, and subsequently diminish the introduced bias with a penalty technique incorporated into the Metropolis algorithm.

The rest of this paper is organized as follows. In Section II, we present our methods, including the RCNN architecture, the training procedure, and the details of the Penalty Ensemble Method. Section III presents the results of our simulations on the $\phi^4$ model, comparing the performance of the PEM against a single RCNN and highlighting the importance of the penalty term. We then discuss the implications of our findings, examining the computational trade-offs and challenging our core assumptions. We conclude with a summary of the paper.

 \section{Metropolis Algorithm}

Our goal is to sample configurations of the lattice $\phi^4$ model, governed by the stationary distribution:

\begin{equation}
p(\mathbf{\phi}) = \frac{e^{-S[\mathbf{\phi}]}}{Z },
\label{boltzman}
\end{equation}
where $Z$ denotes the partition function and $S[\phi]$ denotes the lattice $\phi^4$ action in two dimensions, as defined in \cite{PhysRevD.104.094507}:

\begin{equation}
S[\phi] = \sum_{x \in \Lambda} \Bigl[ -\sum_{\kappa=1}^2 \beta \phi_{x}\phi_{x+e_\kappa} + \phi_{x}^2 + g \left( \phi_{x}^2 - 1 \right)^2 \Bigr],
\label{discreetphi4}
\end{equation}

with $\beta$ denoting the inverse temperature and $g$ the coupling strength. Here, $\Lambda$ denotes the two-dimensional lattice consisting of $L \times L$ sites and $e_{\kappa}$ represents a unit vector in the $\kappa$-th dimension. We impose periodic boundary conditions on the lattice. This model exhibits a phase transition between a paramagnetic and a ferromagnetic phase at a critical point $\beta_c(g)$. We use the $\phi^4$ model to demonstrate our method because of its computational efficiency that allows to isolate and showcase the core functionality of the PEM, even though the method's full potential should be realized in more complex, computationally expensive applications like Density Functional Theory (DFT) calculations for AI-driven molecular dynamics \cite{focassio_performance_2024}.

The Metropolis algorithm constructs a Markov chain by proposing updates to the field and accepting or rejecting them based on the associated change in the action. In the following, we consider a random walk update with Gaussian proposal. The standard steps are as follows:

\begin{enumerate}[noitemsep,nolistsep]
\item Uniformly and randomly select a site $x$ on the lattice.
\item Propose a local update: $\phi_x^\prime = \phi_x + \delta,\quad \delta \sim \mathcal{N}(0,1)$.
\item Compute the action difference $dS(\phi,\phi^\prime) = S[\phi^\prime] - S[\phi]$.
\item Accept the move with probability $p_{\rm acc}(\phi,\phi^\prime) = \min(1, e^{-dS(\phi,\phi^\prime)})$; otherwise, retain the current configuration.
\end{enumerate}

\vspace{.2cm}

Note that since $\phi$ and $\phi^\prime$ differ only at site $x$, we can equivalently write
\begin{equation}
dS(\phi,\phi^\prime) = S_x[\phi_x^\prime] - S_x[\phi_x],
\label{dS_local}
\end{equation}
where
\begin{equation}
S_x[\phi_x] = \phi_x^2 + g(\phi_x^2 - 1)^2 - \beta \phi_x \sum_{\kappa=1}^2 (\phi_{x+e_\kappa} + \phi_{x-e_\kappa}).
\label{local_action}
\end{equation}

These steps are repeated iteratively to construct the Markov chain.

In our work, we aim to emulate this process by approximating the action difference $dS$ using a residual convolutional neural network (RCNN). 

We express $dS$ as a sum of two terms: a gradient term linear in $\delta$, and a nonlinear local term that depends only on $\phi_x$:

\begin{align}
dS(\phi,\phi^\prime) ={}& \, \delta \cdot \frac{\partial S_x}{\partial \phi_x}[\phi_x] 
+ \left(6g\phi_x^2 - 2g + 1\right) \delta^2 \notag\\
&+ 4g\phi_x \, \delta^3 
+ g \, \delta^4 ~.
\label{true_form_ds}
\end{align}

This expression follows from applying a Taylor expansion of $S_x[\phi_x+\delta]$ around $\phi_x$. In particular,
\begin{equation}
S_x[\phi_x+\delta] = S_x[\phi_x] + \sum_{i=1}^\infty \frac{\delta^i}{i!}\frac{\partial^i S_x}{\partial \phi_x^i}[\phi_x].
\label{taylor_expansion}
\end{equation}

Rather than learning $dS$ directly—which can be challenging due to its dependence on the small and fluctuating quantity $\delta$—we focus on learning the gradient of the local action, $\frac{\partial S_x}{\partial \phi_x}[\phi_x]$. This gradient corresponds roughly to the force field, a quantity commonly used in AI-driven molecular dynamics \cite{fu_forces_nodate, chmiela_towards_2018}. The gradient of the action is given by:

\begin{equation}
    \begin{split}
         \frac{\partial S_x}{\partial \phi_x}[\phi_x] \\
        &= -\beta \sum_{\kappa=1}^2 \left( \phi_{x+e_\kappa} + \phi_{x-e_\kappa} \right)
        + 2(1 - 2g)\phi_x + 4g\phi_x^3
    \end{split}
\end{equation}

We train the RCNN to produce an estimate of $\frac{\partial S_x}{\partial\phi_x}[\phi_x]$, which we denote by $\frac{\partial S^{\rm RCNN}_x}{\partial\phi_x}[\phi_x]$. This in turn allows us to estimate $dS(\phi,\phi^\prime)$ by replacing $\frac{\partial S_x}{\partial\phi_x}[\phi_x]$ in Eq.~\eqref{true_form_ds} with $\frac{\partial S^{\rm RCNN}_x}{\partial\phi_x}[\phi_x]$, which we denote by $dS^{\rm RCNN}(\phi,\phi^\prime)$.

Our goal is to evaluate the precision of this prediction—specifically its epistemic uncertainty—and adapt the acceptance probability accordingly to incorporate this uncertainty into the Monte Carlo sampling procedure.

\subsection{Training RCNN}
To train the RCNN (the model's architecture is detailed in Appendix \ref{Appendic B RCNN}), we provide as input a set of independent and equilibrated configurations of the field $\phi$. Each configuration consists of $L \times L$ real-valued components. The target output is the set of gradient components $\nabla S[\phi]$.

For the training procedure, we employ the mean squared error (MSE) loss function. Since we assume that the true gradient components have Gaussian fluctuations, this corresponds to maximizing the likelihood.

After training, the RCNN is able to predict the gradient components of new equilibrated configurations. We denote these predictions as $\nabla S^{\rm RCNN}[\phi]$ and define 

\begin{align}
dS^{\rm RCNN}(\phi,\phi^\prime) ={}& \, \delta \cdot \frac{\partial S^{\rm  RCNN}_x}{\partial \phi_x}[\phi_x] 
+ \left(6g\phi_x^2 - 2g + 1\right) \delta^2 \notag\\
&+ 4g\phi_x \, \delta^3 
+ g \, \delta^4.
\label{rcnn_form_ds}
\end{align}

This prediction of the action difference allows us to perform a new Monte Carlo simulation based on the RCNN prediction and on the acceptance probability:
\begin{equation}
    p_{\rm acc}^{\rm RCNN}(\phi,\phi^\prime) = \min(1, e^{-dS^{\rm RCNN}(\phi,\phi^\prime)}).
    \label{acceptance probability with ds hat}
\end{equation}
We will see that this Monte Carlo simulation based on the RCNN model does not capture the correct macroscopic behavior of the $\phi^4$ model.
To measure the precision of the prediction, we define the residuals as:

\begin{equation}
    R[\phi_x]= \frac{\partial S^{\rm RCNN}_x}{\partial\phi_x}[\phi_x]- \frac{\partial S_x}{\partial\phi_x}[\phi_x].
\end{equation}

Our first goal is to estimate the precision of these predictions, for which we design a simple ensemble method.

\subsection{Penalty Ensemble Method}
\subsubsection{Deep Ensemble}
\label{pem_deep_ensemble}
We adopt a deep ensemble approach \cite{garipov_loss_2018, lakshminarayanan_simple_nodate} to estimate the precision of the predicted gradient components using $N$ independently trained RCNN models, resulting in $N$ distinct predictions. 
As all models are trained on the same dataset, the difference in their predictions stems from the random initialization of their parameters and from the convergence of stochastic optimization algorithms to a local minimum in the model loss. Consequently, each model provides a slightly different estimate of the gradient components:
 
\begin{equation}
    \left\{ \frac{\partial S_x^{\mathrm{RCNN, }(1)}}{\partial \phi_x}[\phi_x], \frac{\partial S_x^{\mathrm{RCNN, }(2)}}{\partial \phi_x}[\phi_x], \ldots, \frac{\partial S_x^{\mathrm{RCNN, }(N)}}{\partial \phi_x}[\phi_x]\right\}.
\end{equation}

For a given component, we define the ensemble mean $\langle \frac{\partial S^{\rm RCNN}_x}{\partial \phi_x}[\phi_x] \rangle_N$ as the average of the $N$ predictions and interpret it as the ensemble's best estimate of $\frac{\partial S_x}{\partial \phi_x}[\phi_x]$. To assess the uncertainty of this estimate, we compute the unbiased sample variance:

\begin{equation}
\begin{aligned}
\sigma^2[\phi_x] &= \frac{1}{N-1} \\
&\quad \times \sum_{i=1}^{N}\Biggl(
\frac{\partial S_x^{\mathrm{RCNN, }(i)}}{\partial \phi_x}[\phi_x]
- \bigl\langle \frac{\partial S_x^{\mathrm{RCNN}}}{\partial \phi_x}[\phi_x] \bigr\rangle_N
\Biggr)^2  .
\end{aligned}
\label{variance first estimator}
\end{equation}

Assuming the predictions are unbiased (i.e., the residuals have zero mean) and possess finite variance, the Central Limit Theorem implies that, as \(N \rightarrow \infty\), the ensemble mean converges to $\frac{\partial S_x}{\partial \phi_x}[\phi_x]$ with Gaussian fluctuations of order $\sigma[\phi_x] / \sqrt{N}$.
\subsubsection{Noise Penalty Method}

The noise penalty method~\cite{Ceperley_1999} addresses the challenge of sampling from
distributions where the energy cannot be computed exactly but is instead estimated
with statistical noise. In this framework, the acceptance probability is modified by
including a penalty term that ensures detailed balance is satisfied on average, leading
to statistically correct sampling.

Following the previous discussion, we introduce the notation
$\left\langle dS^{{\rm RCNN}}(\phi,\phi')\right\rangle_N$
to denote the estimate of $dS(\phi,\phi^\prime)$ obtained by replacing
$\frac{\partial S_x}{\partial\phi_x}[\phi_x]$ in Eq.~\eqref{true_form_ds}
with the ensemble-averaged gradient
$\left\langle \frac{\partial S^{\rm RCNN}_x}{\partial \phi_x}[\phi_x] \right\rangle_N$.

In our setting, using the ensemble approach to estimate the noise of the gradient in
Eq.~\eqref{true_form_ds}, the acceptance probability resulting from the noise penalty
method reads
\begin{equation}
\begin{aligned}
p_{\text{acc}}^{\text{PEM}}(\phi,\phi')
&= \min\left(1, e^{-\Delta_{\rm PEM}(\phi,\phi')}\right), \\
\Delta_{\rm PEM}(\phi,\phi')
&= \left\langle dS^{{\rm RCNN}}(\phi,\phi')\right\rangle_N
+ \frac{\sigma^2[\phi_x]\,\delta^2}{2N}.
\end{aligned}
\label{expanded penalty}
\end{equation}
where  $\phi$ and $\phi'$ are two configurations that differ only at site $x$.
The quantity $\sigma^2[\phi_x]$ denotes the variance of the local gradient estimator and
depends only on the local field value $\phi_x$. For details on the derivation of this expression, we refer the reader to Appendix~B.

\section{Results for Metropolis simulations}

\subsection{Ferromagnetic and Paramagnetic Phases}

\begin{figure}[ht!]
    \centering
    \includegraphics[width=1.0\linewidth]{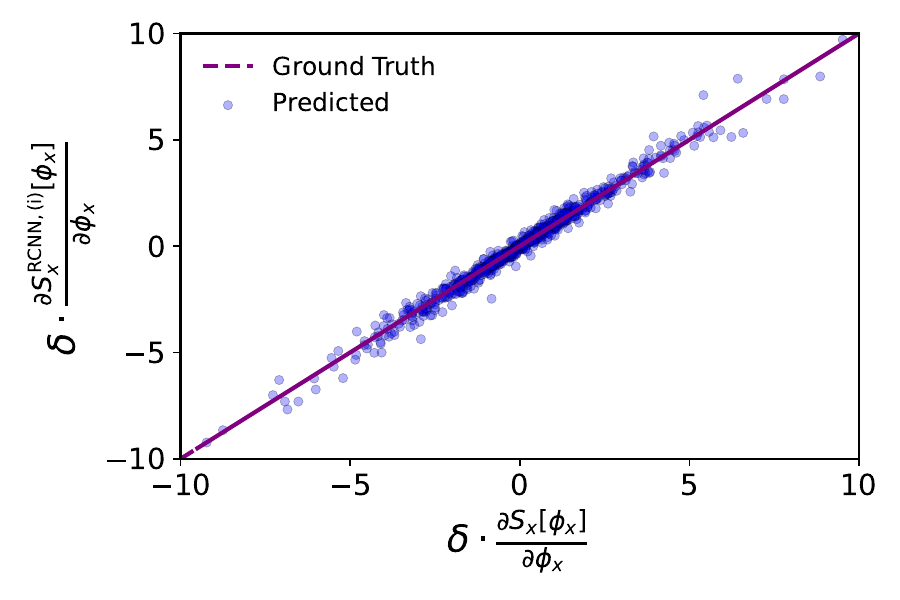}
    \caption{Performance of the RCNN prediction on test data for the ferromagnetic phase $\beta=0.7$ for $1000$ samples.}
    \label{performance_grad}
\end{figure}

\captionsetup{margin=0pt}
\begin{figure*}[ht!]
    \centering
    \captionsetup{justification=raggedright, singlelinecheck=false, format=plain}

    \begin{subfigure}[t]{0.48\textwidth}
        \includegraphics[width=\linewidth]{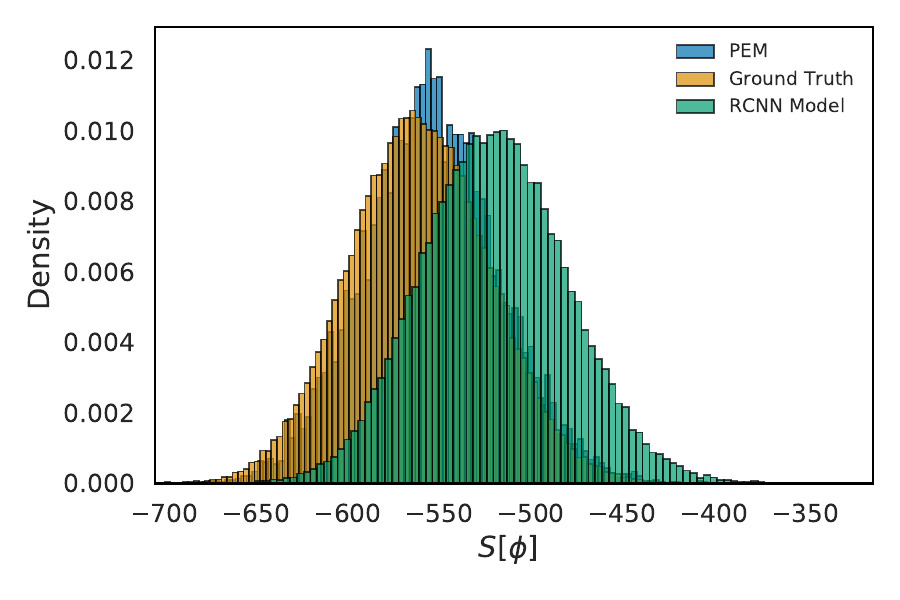}

        \label{fig:mh_ferro}
    \end{subfigure}
    \hfill
    \begin{subfigure}[t]{0.48\textwidth}
        \includegraphics[width=\linewidth]{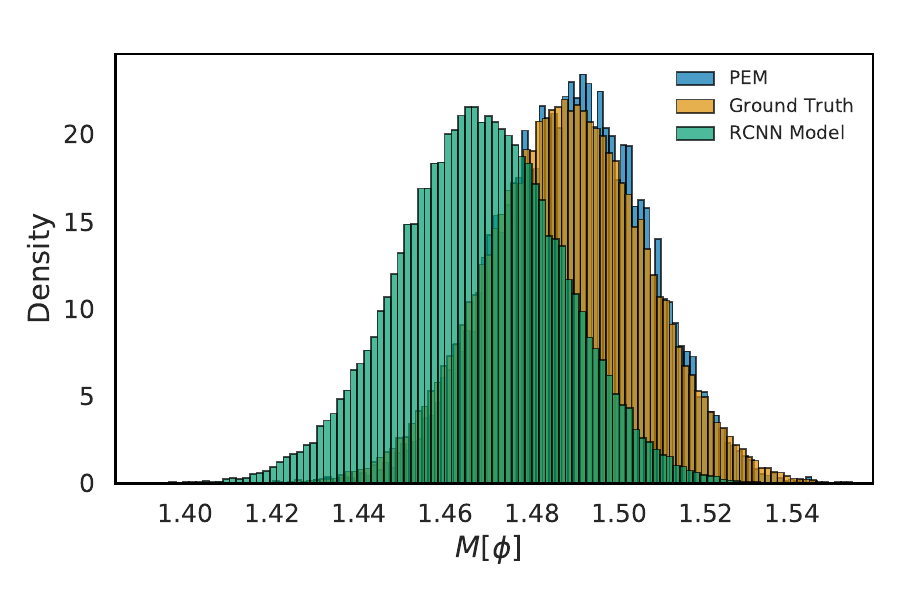}

        \label{fig:mala_ferro}
    \end{subfigure}
    \hfill
    \begin{subfigure}[t]{0.48\textwidth}
        \includegraphics[width=\linewidth]{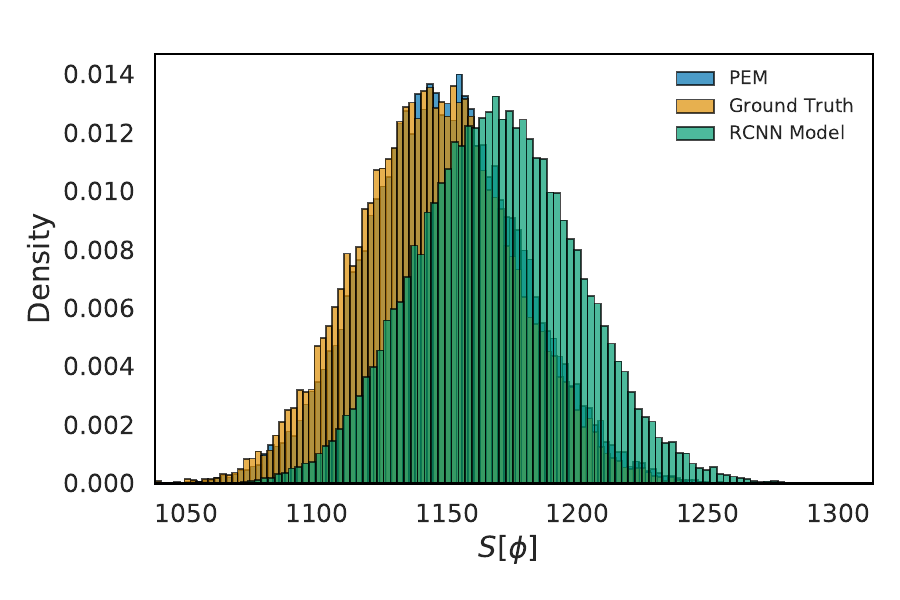}

        \label{fig:mh_param}
    \end{subfigure}
    \hfill
    \begin{subfigure}[t]{0.48\textwidth}
         \includegraphics[width=\linewidth]{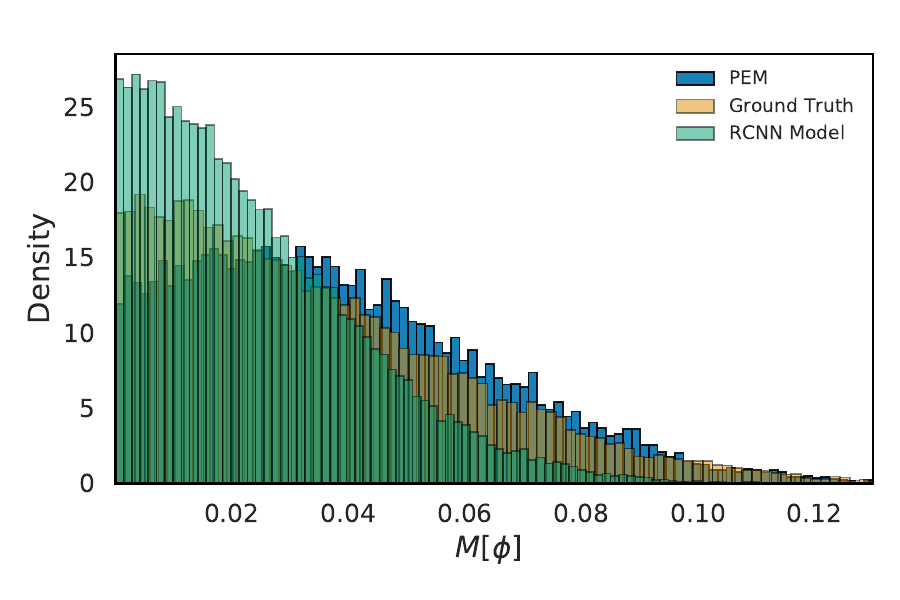}

        \label{fig:mh_algo}
    \end{subfigure}

    \caption{Density of $1000$ equilibrated, independent configurations of size $48 \times 48$ for the $\phi^4$ model. The RCNN (green) does not sample the correct configurations, while the  PEM with $N=3$ (blue) corrects the bias. Upper panels: Ferromagnetic Phase for $\beta=0.7$. Lower panels: Paramagnetic Phase for $\beta=0.5$. We collected samples by running 20 independent Markov chains and recording one sample every $ 46080$ steps.}
    \label{fig:action_density_grid}
\end{figure*}

We perform Monte Carlo simulations of $48\times 48$ configurations of the $\phi^4$ model, using $g=0.1$. For this model, the inverse critical temperature is expected at $\beta_{c} \simeq 0.6$. We first simulate at inverse temperature $\beta = 0.7$, corresponding to the ferromagnetic phase, and at $\beta = 0.5$, corresponding to the paramagnetic phase. Standard Monte Carlo sampling is initially performed using the exact evaluation of the gradient of the action (see Eq.~\eqref{true_form_ds}). The data obtained from this simulation, referred to as the Ground Truth, serve as a benchmark for the data generated using the RCNN architecture, and also provide the configurations used to train the RCNN model. In all simulations we initialized the sampling with equilibrated configurations.

The Monte Carlo computation using the RCNN model does not sample the target
distribution if one initializes the Markov chain from a non-equilibrated configuration
(with or without PEM). This behavior is expected, as neural networks are known to provide
poor predictions when presented with atypical configurations lying outside their
training distribution \cite{YUAN2022115569}.

Then, we perform additional Monte Carlo simulations using the same parameters, but now incorporating the RCNN model predictions into the acceptance probability, as defined in Eq.~(\ref{acceptance probability with ds hat}). Each RCNN model is trained on the same 90 independent and equilibrated configurations previously sampled.
We intentionally limit the training set size to reflect the data scarcity typically encountered in realistic physical systems. For optimization, we use the Adam algorithm, training for 150 epochs in the ferromagnetic phase and 180 epochs in the paramagnetic phase, with a batch size of 32. In Fig.~\ref{performance_grad}, we assess the performance of the RCNN model by testing it on a set of configurations sampled from the ground truth algorithm. We observe that the RCNN predictions are consistently accurate across these configurations.

In Fig.~\ref{fig:action_density_grid}, we present densities of the estimated values of the action and magnetization, defined as $M[\phi]=\left| \frac1{L^2}\sum_{x\in \Lambda}\phi_x \right|$ for configurations sampled using the two different Monte Carlo methods. Although the gradient prediction errors are small, the configurations generated by the RCNN-based Monte Carlo still show significant deviations from those obtained with the ground truth method, in both the ferromagnetic and paramagnetic phases. To correct these deviations, we implement the PEM described in the previous section. In the following, all experiments use an ensemble of $N = 3$ independently trained models. We provide additional results for different values of $N$ in the Appendix~\ref{appendix: observable for 5 models} that exhibit a similar behavior to $N=3$. In the Metropolis algorithm, for each randomly selected gradient component, we compute the ensemble mean $\langle \frac{\partial S^{\rm RCNN}_x}{\partial \phi_x}[\phi_x] \rangle_N$ and variance $\sigma^2[\phi_x]$.
The acceptance probability is then modified by incorporating the penalty factor, as defined in Eq.~\eqref{expanded penalty}. Our results, shown in Fig.~\ref{fig:action_density_grid}, demonstrate that the PEM successfully reproduces the ground truth distributions.

\subsection{Close to Criticality}
\label{close_to_criticality}
We repeat this study near the transition point by sampling equilibrated, independent configurations at the inverse temperature $\beta =0.6$, which we estimate to be close to criticality. Near the critical region, fluctuations become stronger and longer-ranged. To address this, we increase the RCNN training set size by a factor of 10 and perform gradient descent for 200 epochs with a batch size of 32.

\begin{figure}[ht!]
\centering
\captionsetup{justification=raggedright, singlelinecheck=false, format=plain}

\includegraphics[width=1.02\linewidth]{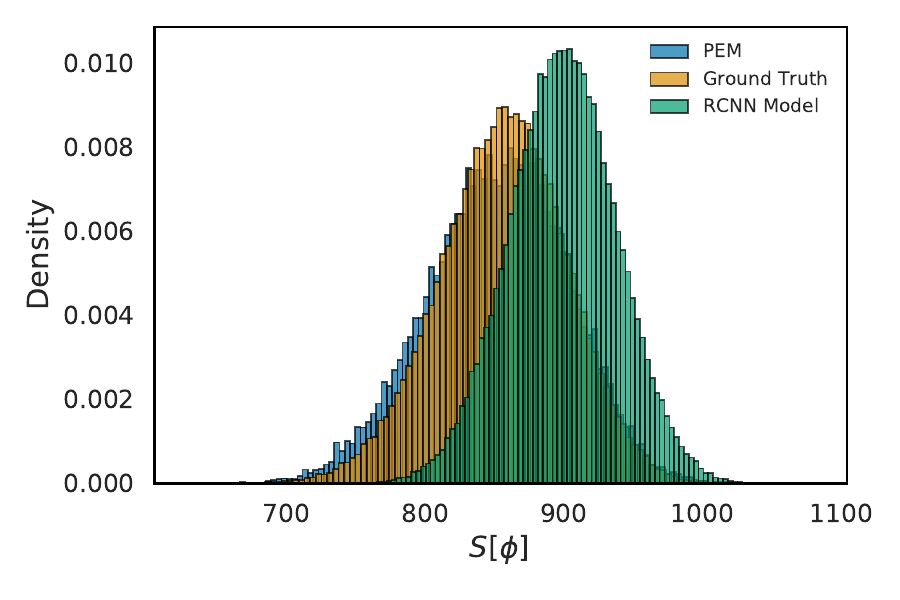}
\includegraphics[width=1.02\linewidth]{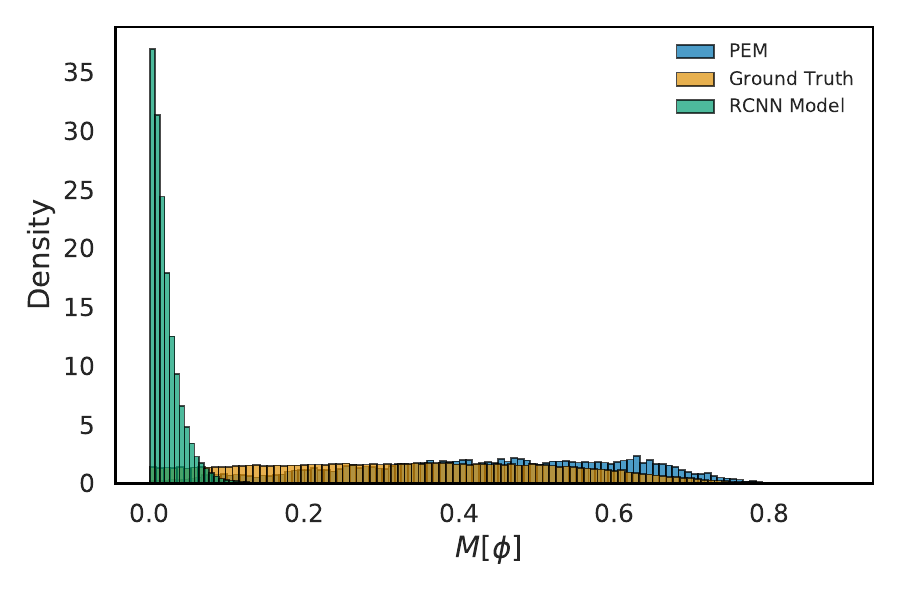}
\caption{Densities of 500 equilibrated, independent configurations of size $48 \times 48$ at the critical point $\beta_{c}=0.6$ for the $\phi^4$ model. The RCNN (green)  samples paramagnetic configurations, while the PEM with $N=3$ (blue)  recovers the correct behavior of the critical point. We collected samples by running 20 independent Markov chains and recording one sample every $92160$ steps.}
\label{phase_transition}
\end{figure}

In Fig.~\ref{phase_transition}, we observe that the Metropolis algorithm using only the RCNN model tends to sample configurations biased toward the paramagnetic phase. Specifically, the magnetization density peaks around zero while the action is pushed towards higher values with respect to
the correct one. A possible explanation for this behavior is that the residual errors of the RCNN act as disorder that pushes the system toward the paramagnetic phase.

Next, we apply the PEM to sample configurations from the critical region. The results demonstrate that the PEM is able to better capture the actual behavior of the system, even at the critical point.
\captionsetup{margin=0pt}

\subsection{Non Gaussian tail in the residual distribution}

\begin{figure}[ht!]
\centering
\captionsetup{justification=raggedright, singlelinecheck=false, format=plain}

\includegraphics[width=1.\linewidth]{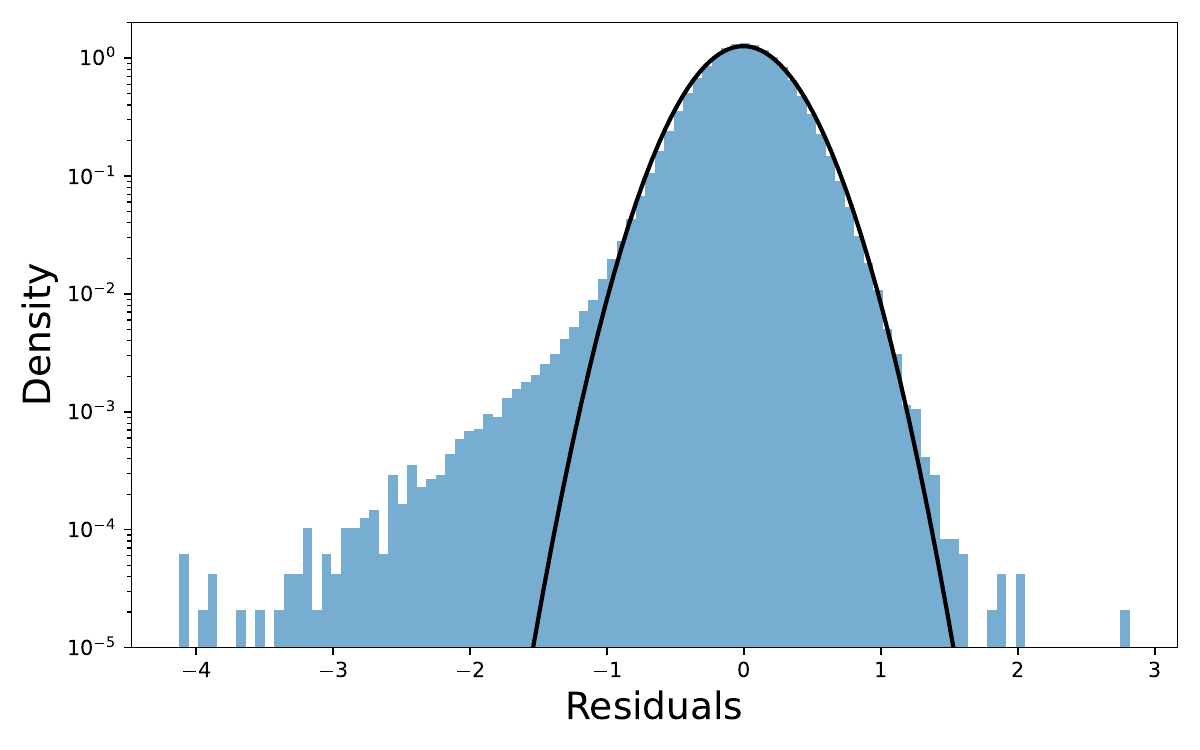}

\includegraphics[width=1.\linewidth]{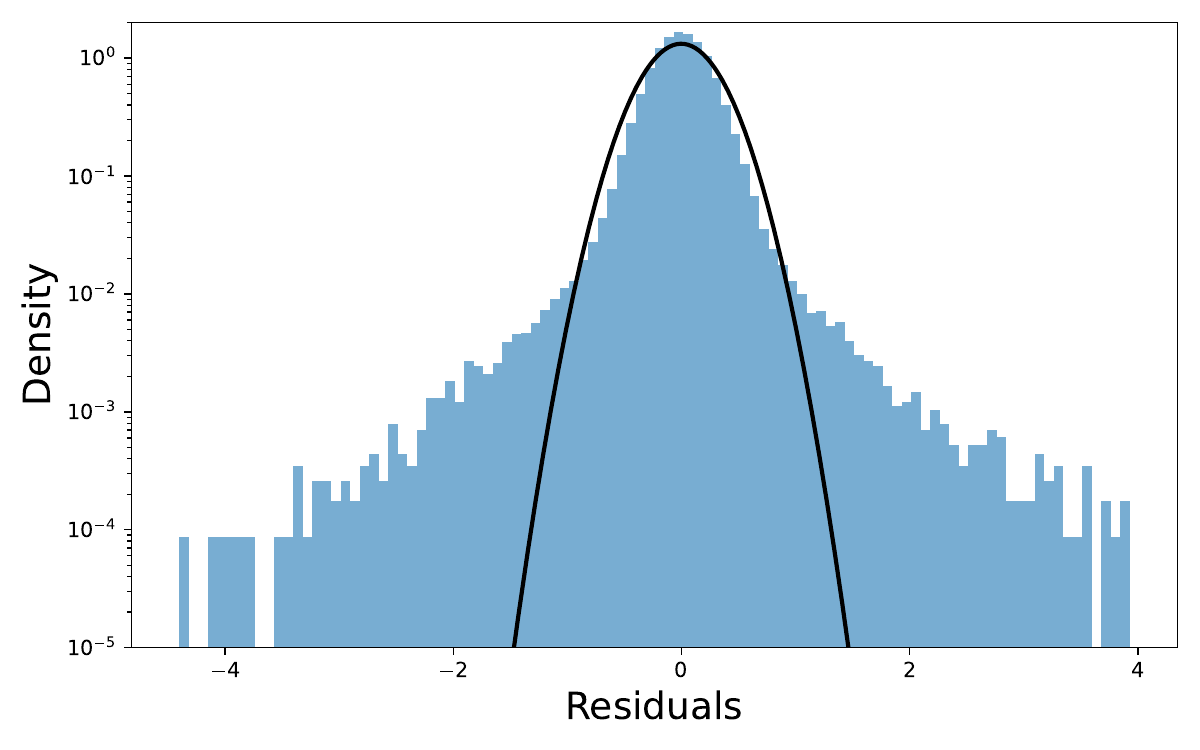}
\caption{
    Density of the residuals. 
    Upper panel: Ferromagnetic phase $\beta = 0.7$. Solid black line is the Gaussian fit with $\mu = 0.$ and $\sigma = 0.3$. 
    Lower panel: Paramagnetic phase $\beta =0.5$. Solid black line is the Gaussian fit with $\mu = 0.$ and $\sigma = 0.3$. For these figures we have retrieved all the gradients of all lattice sites.
    }
\label{fitted_gaussian}
\end{figure}

In this part, we verify the applicability of the penalty method. Specifically, we test the assumption that the residuals of the gradient components can be modeled as centered Gaussian random variables.
In Fig.~\ref{fitted_gaussian}, we show histograms of the residuals for both the ferromagnetic and paramagnetic phases. On a semilog scale, the bulk of the distribution closely follows a Gaussian profile, supporting the applicability of the penalty method. However, deviations from Gaussianity appear in the tails below approximately $\sim 10^{-2}$. Notably, a left-sided exponential tail is observed at the ferromagnetic point, while a symmetric exponential tail emerges at the paramagnetic point. These discrepancies likely stem from the RCNN’s limited capacity to generalize in sparsely sampled regions of configuration space, where rare force configurations occur. While the tail behavior does not adhere to that of a Gaussian, the impact on the simulation seems limited. This likely stems from the fact that the tails only account for a small part of the entire distribution. Additionally, large deviations are associated with high predictive uncertainty and are therefore exponentially likely to be rejected, which helps to mitigate their effect on the simulation. Further details on this effect can be found in Appendix \ref{appendix of gradients}.

\begin{figure}[ht!]
\centering
\captionsetup{justification=raggedright}
\includegraphics[width=1.0\linewidth]{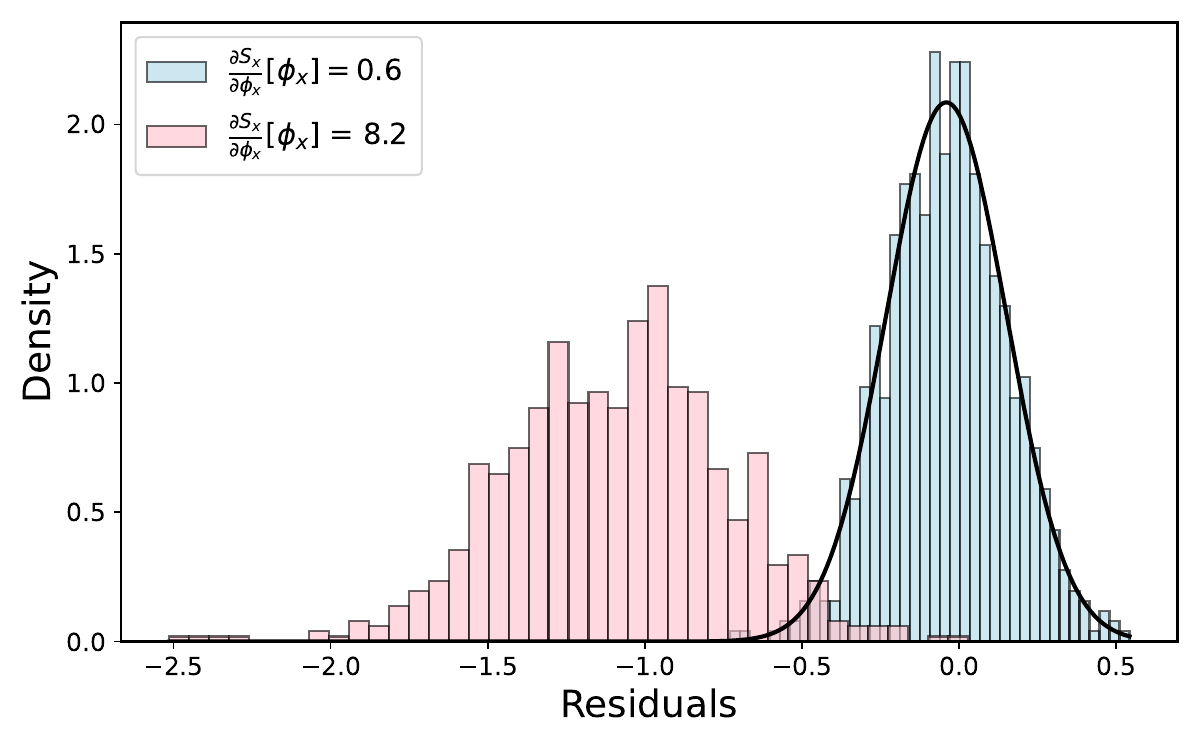}
\caption{Density of residuals for a given configuration and gradient component. The configuration is equilibrated in the ferromagnetic phase and the histogram is derived from 800 independently trained models. The light blue density represents a typical gradient value $\left( \frac{\partial S_x}{\partial \phi_x}[\phi_x] = 0.6 \right)$. Solid black line represents the Gaussian fit with $\mu = 0.027$ and $\sigma=0.3$. The pink density represents an atypical gradient value $\left(\frac{\partial S_x}{\partial \phi_x}[\phi_x] = 8.2 \right)$.  } 
\label{gaussian_800}
\end{figure}

To better understand the RCNN's limitations in underrepresented regions, we conduct the following experiment. We first identify a configuration in the ferromagnetic phase containing an atypical gradient component with a high-magnitude value (specifically, $\frac{\partial S_x}{\partial \phi_x}[\phi_x] = 8.2$). An ensemble of 800 RCNN models is then trained, and we collect their predictions for this specific component. For comparison, we select a typical component ($\frac{\partial S_x}{\partial \phi_x}[\phi_x] = 0.6$) and evaluate its predictions using the same trained models. The resulting residual densities are shown in Fig.~\ref{gaussian_800}. They reveal that predictions for the typical component closely follow a Gaussian distribution, whereas the residuals for the rare component are clearly biased and exhibit heavier tails, indicating the model’s difficulty in capturing rare force configurations accurately.

\subsection{Runtime performance and acceptance probabilities}

\begin{figure}[h!]
\centering
\captionsetup{justification=raggedright, singlelinecheck=false, format=plain}

\includegraphics[width=1.02\linewidth]{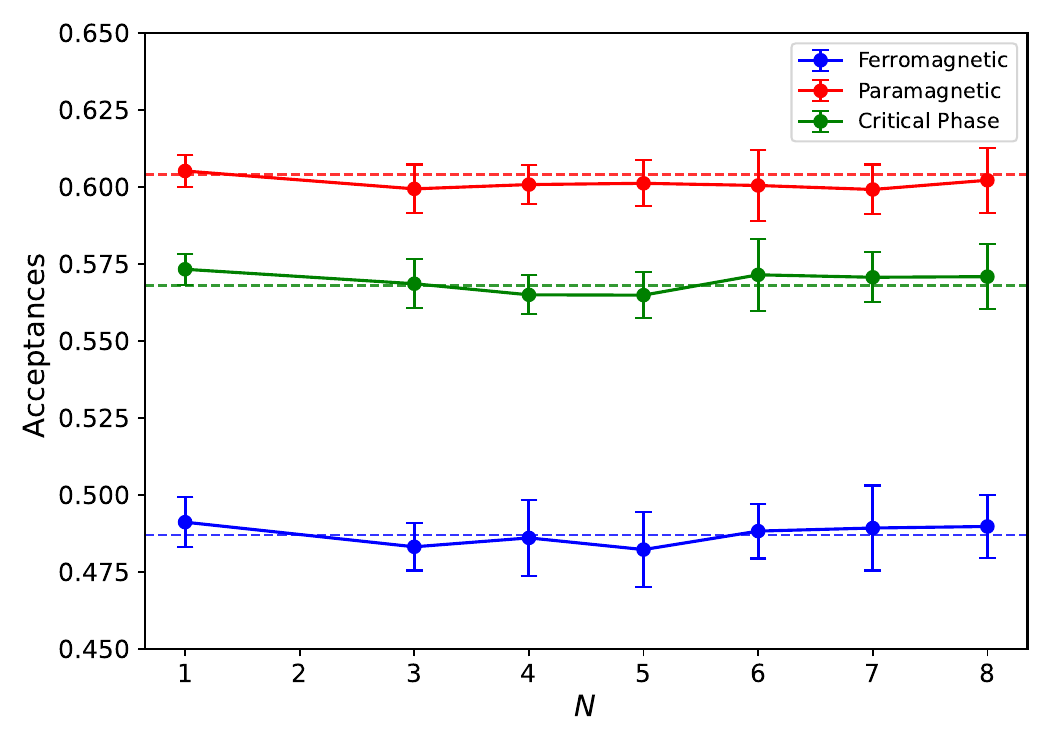}

\caption{Average PEM acceptance ratios across the three phases as a function of the ensemble size $N$. The case $N=1$ represents the baseline RCNN Monte Carlo simulation without PEM. Dashed horizontal lines indicate the ground truth reference values for the average acceptance ratio. The error bars represent the standard deviation of the mean acceptance, estimated from 10 independent Markov chains, each run for 2304 iterations.
}
\label{acceptances}
\end{figure}

The use of a neural network ensemble necessarily introduces a computational overhead: the algorithm's runtime scales with the number of ensemble members, $N$, due to the repeated evaluation of $N$ networks during the Monte Carlo procedure. This is compounded by the one-time, up-front cost of training multiple models. Regarding acceptance probabilities, we find that both the single RCNN and the PEM yield high acceptance ratios, comparable to those achieved with the exact action calculation. For the chosen size of the proposal step $(\delta)$ in our Metropolis algorithm, the ground truth acceptance probabilities are 48.7\% (ferromagnetic), 60.4\% (paramagnetic), and 56.8\% (phase transition). The PEM introduces a small and acceptable decrease in this ratio, specifically a reduction of only 0.5\% to 1.5\% across all test phases as shown in Fig.~\ref{acceptances}. In principle, one might expect the performance of PEM to improve with the number of models $N$, as suggested in Section~\ref{pem_deep_ensemble} (deep ensemble) through a central-limit-type argument. In practice, however, increasing $N$ beyond a moderate value does not enhance sampling performance (details in Appendix~\ref{appendix: observable for 5 models}). Although a larger ensemble reduces the noise penalty and therefore increases the acceptance rate, it does not translate into better exploration. The reason is that the method relies on the assumption of unbiased model predictions; while the variance decreases with $N$, any systematic bias remains constant. For sufficiently large $N$, this residual bias becomes the dominant source of error, limiting the gains from additional models.

\captionsetup{margin=0pt}

\begin{figure}[ht!]
\centering
\captionsetup{justification=raggedright, singlelinecheck=false, format=plain}

\includegraphics[width=1.02\linewidth]{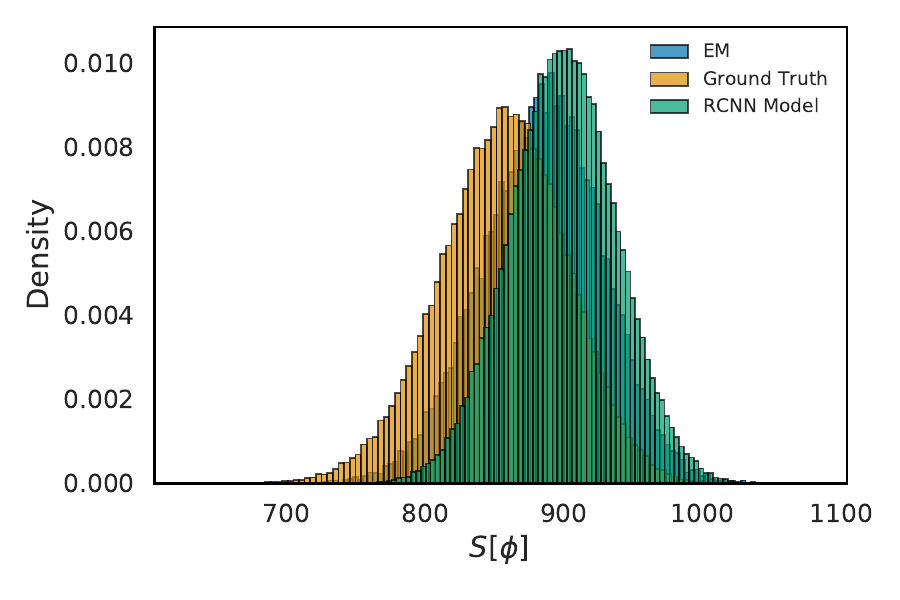}
\includegraphics[width=1.02\linewidth]{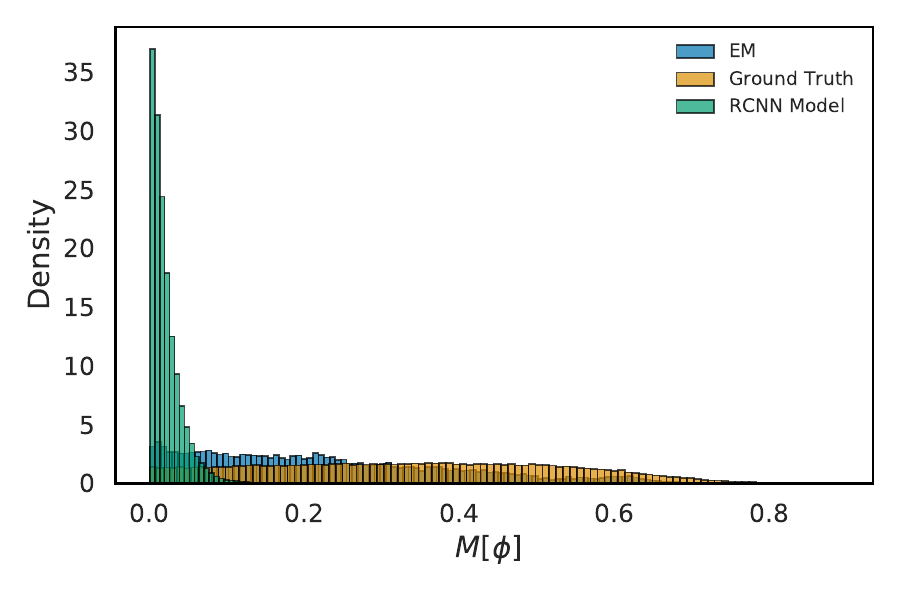}

\caption{Densities of 500 equilibrated, independent configurations of size $48 \times 48$ at the critical point $\beta_{c}=0.6$ for the $\phi^4$ model. While the RCNN (green) incorrectly samples paramagnetic configurations, the Ensemble Method (blue) successfully captures the critical behavior, though its performance remains significantly inferior to the PEM. We collected samples by running 20 independent Markov chains and recording one sample every $92160$ steps.}
\label{phase_transition_wp}
\end{figure}

In our experiments, the value $N=3$ provides the best trade-off between reliable variance estimation and computational efficiency. Fig.~\ref{phase_transition_wp} illustrates the contribution of the ensemble alone, by comparing the Ensemble Method (EM) -- which averages the predictions of $N=3$ RCNNs without including the noise penalty -- to a single RCNN. While the deep ensemble alone already reduces the variance of the estimator given by Eq.~\eqref{expanded penalty} and partially mitigates the bias, it does not fully recover the ground truth distribution. This shows that the noise penalty factor remains necessary to account for the residual variance of this estimator, as demonstrated by the full PEM results in Fig.~\ref{phase_transition}.

\captionsetup{margin=0pt}

\begin{figure}[h!]
    \centering
    \includegraphics[width=1.\linewidth]{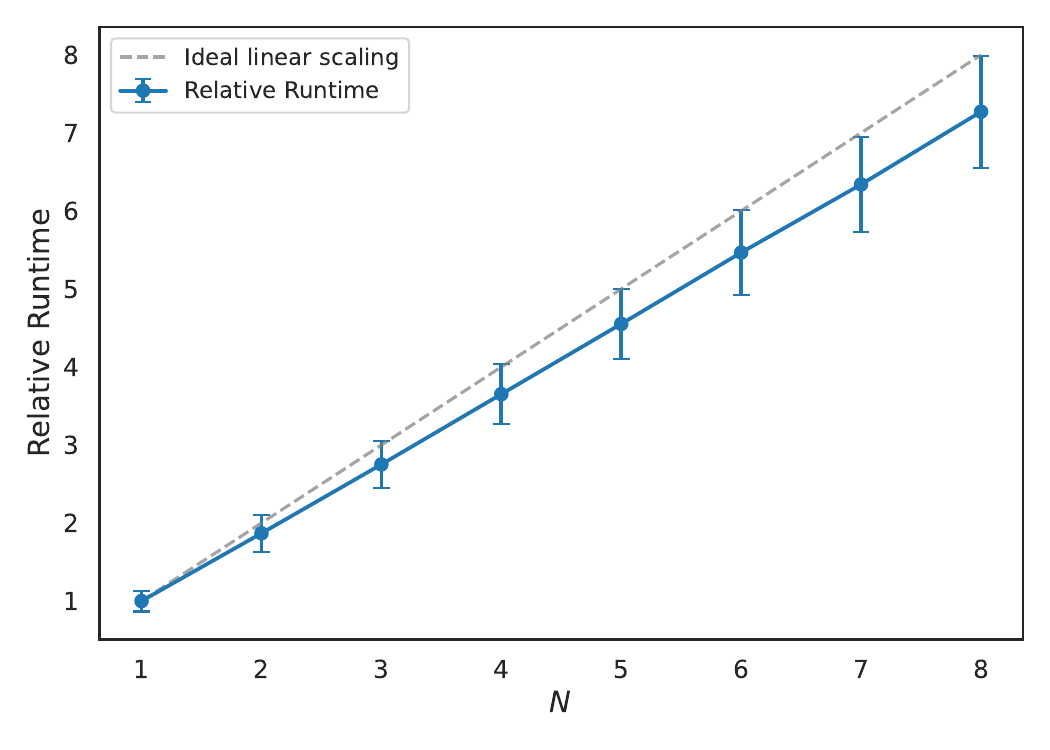}
    \caption{Relative runtime of the PEM over a single RCNN model ($N=1$), plotted as a function of the ensemble size $N$. The error bars represent one standard deviation around the mean of the relative runtime, estimated from 10 independent Markov chains, each run for 2304 iterations.}
    \label{fig:rel_runtime}
\end{figure}

Overall, the runtime of PEM is larger than that of a single RCNN model, due to three factors: (i) the training of multiple models, (ii) the slight reduction in acceptance ratios, and (iii) the repeated use of $N$ models within the Monte Carlo procedure. In Fig.~\ref{fig:rel_runtime}, we report the relative runtime of PEM with respect to a single RCNN ($N=1$), as a function of the ensemble size $N.$ 

Finally, we consider whether increasing the representational capacity and the computational cost of a single model might outperform the PEM. To evaluate this, Appendix~\ref{appendix of advanced model} provides a comparative analysis using an augmented RCNN architecture with significantly increased depth and parameterization. Despite its increased expressivity, this model exhibits only marginal performance improvements and does not match the robustness achieved by PEM. This suggests that the primary bottleneck in sampling fidelity is not the model’s approximation power, but rather the absence of a mechanism to account for epistemic uncertainty.

\section{Conclusion}

In this work, we investigated the use of RCNN-based gradient estimators in Monte Carlo sampling. The predictive noise inherent to the RCNN naturally introduces stochasticity, embedding a form of epistemic uncertainty into the sampling dynamics and thus introducing a bias in macroscopic observables. Our empirical analysis, using the two-dimensional $\phi^4$ model and an RCNN-based gradient estimator within the Metropolis algorithm, clearly demonstrated that even a well-performing surrogate model can introduce significant bias into the sampling process, leading to inaccurate estimates of critical physical observables.

To mitigate this issue, we proposed the Penalty Ensemble Method, which quantifies the predictive variance through an ensemble technique. This uncertainty information is then integrated into the Monte Carlo sampling through a noise penalty that modifies the Metropolis acceptance rule under the assumption of Gaussian-distributed predictions. By selectively increasing the rejection probability in regions of high predictive uncertainty, PEM effectively discourages the sampling of out-of-distribution states. Our results demonstrate that PEM successfully diminishes the introduced bias. Furthermore, the observed robustness of PEM, even when residuals exhibit heavy-tailed behavior, highlights its practical applicability beyond idealized Gaussian noise assumptions.

The successful implementation of PEM represents a significant step towards more robust and reliable integration of deep learning into scientific computing, addressing a major bottleneck for the widespread application of AI in complex stochastic systems. While PEM effectively mitigates bias, future work could explore refinements such as more advanced uncertainty quantification techniques beyond empirical variance, especially when dealing with highly non-Gaussian error distributions. Additionally, instead of the standard MSE loss function, a more physically informed loss function might help to mitigate this effect. Finally, the penalty mechanism we introduced can be naturally extended to more general sampling schemes, such as Langevin or Smart Monte Carlo \cite{rossky_brownian_1978}, opening up further applications in molecular simulation.

\textit{Acknowledgments.} We thank Markus Holzmann for seminal
discussions and Élinor Berger for useful comments.

\FloatBarrier 
\bibliographystyle{apsrev4-2}  
\bibliography{apssamp}   

\appendix
\clearpage
\section{Residual Convolutional Neural Networks approximation.}
\label{Appendic B RCNN}

To approximate $\nabla S$ in $\phi^4$ field theory, Convolutional Neural Networks (CNNs) provide an effective regression tool. Their main advantage in lattice field theory is the natural ability to capture translational invariance, a fundamental symmetry of the system. A crucial technical aspect is the use of periodic boundary conditions during training, implemented via periodic padding in the RCNN architecture. This ensures consistent interactions between neighboring lattice sites, even across boundaries. In contrast, simple feedforward neural networks often introduce strong biases, resulting in poor predictive performance. Our RCNN architecture, illustrated in Fig.~\ref{fig:res_cnn}, begins with a convolutional layer, followed by batch normalization and a ReLU activation. This is followed by a residual block composed of another convolutional layer, batch normalization, and ReLU activation. To align feature dimensions, a projection layer is included in the residual connection. The output of the residual block is then summed with the projected connection and passed through a final convolutional layer to produce the output. The residual connection allows the network to bypass transformations when needed, mitigating vanishing-gradient issues and improving feature propagation. This enables deeper feature extraction while retaining important information from earlier layers. For training, we used the ADAM optimizer with a fixed learning rate of $\eta = 0.001$.

\vspace{.2cm}

\begin{figure}[h!]
    \centering
    \begin{tikzpicture}[
        layer/.style={rectangle, draw, minimum width=1.5cm, minimum height=0.8cm, text centered, font=\small},
        input/.style={layer, fill=yellow!30},
        conv/.style={layer, fill=orange!30},
        bn/.style={layer, fill=blue!30},
        relu/.style={layer, fill=red!30},
        res/.style={layer, fill=gray!30},
        output/.style={layer, fill=green!30},
        arrow/.style={->, thick}
    ]
    
    % Input
    \node[input, fill=yellow!30] (input) {Input, $\phi$, configuration};
    % \node[layer] (input) {Input, $\vec{\phi}$, configuration};
    
    % Initial Conv Block
    \node[conv, below=0.4cm of input] (conv1) {Conv2D (1, 4)};
    \node[bn, right=0.4cm of conv1] (bn1) {BatchNorm2D};
    \node[relu, right=0.4cm of bn1] (relu1) {ReLU};
    
    % Residual Block
    \node[conv, below=0.6cm of conv1] (conv2) {Conv2D (4, 8)};
    \node[bn, right=0.4cm of conv2] (bn2) {BatchNorm2D};
    \node[relu, right=0.4cm of bn2] (relu2) {ReLU};
    
    % Residual Connection
    \node[res, below=0.6cm of conv2, minimum width=1cm] (residual) {Projection (4, 8)};
    
    % Output Conv Block
    \node[conv, below=0.6cm of residual] (output_conv) {Conv2D (8, 1)};
    \node[output, below=0.4cm of output_conv] (output) {Output,$\nabla  S^{\rm RCNN}[\phi]$};
    
    % Arrows
    \draw[arrow] (input) -- (conv1);
    \draw[arrow] (conv1) -- (bn1);
    \draw[arrow] (bn1) -- (relu1);
    \draw[arrow] (relu1) -- (conv2);
    \draw[arrow] (conv2) -- (bn2);
    \draw[arrow] (bn2) -- (relu2);
    \draw[arrow] (relu2) -- (residual);
    \draw[arrow] (residual) -- (output_conv);
    \draw[arrow] (output_conv) -- (output);
    
    % Residual connection
    \draw[arrow] (relu1.east) --++ (0.2,0) |- (residual.east);
    
    \end{tikzpicture}
    \caption{RCNN architecture with residual blocks.}
    \label{fig:res_cnn}
\end{figure}

\section{Noise Penalty Method Heuristics.}
\label{appendix: noise penalty}

In this section, we provide an informal derivation of the noise penalty method as given
by Ceperley and Dewing~\cite{Ceperley_1999}. Note that in the original derivation, it is
assumed that when a move from $\phi$ to $\phi'$ is proposed, the estimate of the energy
difference has a Gaussian error with zero mean and a standard deviation $\sigma$,
independent of $\phi$ and $\phi'$.
The same result, however, holds in the more general
case where the standard deviation depends on $\phi$ and $\phi'$, i.e., $\sigma = \sigma(\phi,\phi')$. In our implementation, the standard deviation
is estimated empirically and depends on $\phi$.

Our goal is to derive the modified acceptance probability
$p^{\rm PEM}_{\text{acc}}(\phi,\phi')$, based on the RCNN estimate
$dS^{\text{RCNN}}$ of the true energy difference $dS$. This probability must satisfy the
detailed balance condition \textbf{on average}:
\begin{equation}
\label{eq:avg_detailed_balance}
\left\langle p^{\rm PEM}_{\text{acc}}(\phi',\phi) \right\rangle
=
e^{-dS}
\left\langle p^{\rm PEM}_{\text{acc}}(\phi,\phi') \right\rangle .
\end{equation}
Here, the ensemble average $\langle \ldots \rangle$ is taken over all possible estimates
$dS^{\text{RCNN}}$. For a given transition $\phi \to \phi'$, this corresponds to averaging
over the distribution of $dS^{\text{RCNN}}$ entering
$p^{\rm PEM}_{\text{acc}}(\phi,\phi')$.

We assume that these estimates have Gaussian fluctuations around the true value $dS$,
namely:
\begin{equation}
dS^{\text{RCNN}} = dS + \sigma(\phi,\phi')\,\epsilon ,
\label{eq:gaussian_average}
\end{equation}
where $\epsilon$ is a normally distributed random variable.

A solution to Eq.~\eqref{eq:avg_detailed_balance} is given by
\begin{equation}
p^{\rm PEM}_{\text{acc}}(\phi,\phi')
=
\min\left(1, e^{-\Delta_{\rm PEM}(\phi,\phi')}\right),
\label{eq:penalty_method}
\end{equation}
where the noise-penalty exponent is defined as
\begin{equation}
\Delta_{\rm PEM}(\phi,\phi')
=
\left\langle dS^{{\rm RCNN}}(\phi,\phi')\right\rangle_N
+
\frac{1}{2}\,\sigma^2(\phi,\phi').
\label{eq:Delta_general}
\end{equation}
Eq.~\eqref{eq:penalty_method} should be averaged over all possible estimates
$dS^{\text{RCNN}}$ assuming Eq.~\eqref{eq:gaussian_average}. It can then be verified that
this expression ensures the detailed balance condition~\eqref{eq:avg_detailed_balance}
on average.

In our setting, the variances of the residuals entering the stochastic estimate of the
gradient are not known \emph{a priori}. We therefore rely on the empirical estimate
$\sigma^2[\phi_x]$ defined in Eq.~\eqref{variance first estimator}, which depends only on
the local field value $\phi_x$. This corresponds to the following specialization of
$\Delta_{\rm PEM}(\phi,\phi')$:
\begin{equation}
\Delta_{\rm PEM}(\phi,\phi')
=
\left\langle dS^{{\rm RCNN}}(\phi,\phi')\right\rangle_N
+
\frac{\sigma^2[\phi_x]\,\delta^2}{2N},
\end{equation}
which leads to the acceptance probability employed in the main text, Eq.~\eqref{expanded penalty}.

While it is in principle possible to further correct for the uncertainty introduced by the
estimation of the variance~\cite{Ceperley_1999}, we do not consider such refinements here.

% \section{Proof of the variational action of the $\phi^{4} $ model.}
% One can determine the variation of the action through the variation of the field $\phi_x \rightarrow \phi_x + \delta \phi$ where $x \in  (L)$ where $L$ denotes to Lattice, representing the point $(i,j)$, starting with the formula of Eq.\ref{discreetphi4}. So we have:
% \begin{widetext}
%     \begin{equation}
%     \begin{split}
%         S[\phi + \delta \phi] - S[\phi] = \sum_{x \in L} \Bigl\{ -\sum_{\kappa=1}^{2} \beta(\phi_{x+e_{\kappa}}&(\phi_{x} + \delta \phi)) +  \left(\phi_{x}+\delta \phi \right)^2+g \left( (\phi_{x}+\delta\phi)^2 -1 \right)^2 \Bigl\} \\
%         \vspace{1cm}
%         - \sum_{x \in L}& \Bigl\{ -\sum_{\kappa=1}^2 \beta \phi_{x}\phi_{x+e_\kappa}  + \phi_{x}^2 +g \left( \phi_{x}^2 -1 \right)^2 \Bigr\}\\
%         \vspace{1cm}
%         =  \sum_{x \in L} \Bigl\{ -\sum_{\kappa=1}^{2} 2\beta( \phi_{x+e_{\kappa}} \delta) &+  \left(2\phi_{x}\delta\phi+\delta\phi^2\right)(1-2g)+g \left( (\phi_{x}+\delta\phi)^4 - \phi_{x}\right)^4 \Bigl\} \Rightarrow\\
%         \vspace{1cm}
%         dS_{analytical} = \sum_{x \in L} \Bigl\{ -\sum_{\kappa=1}^{2} 2\beta(\phi_{x+e_{\kappa}}\delta\phi) &+  \left(2\phi_{x}\delta\phi+\delta\phi^2\right)(1-2g)+g \left( (\phi_{x}+\delta\phi)^4 - \phi_{x}\right)^4 \Bigl\}
%     \end{split}
% \end{equation}
% \end{widetext}

\section{PEM observables for $N=4,\ 5,\ 6,\ 7,\ 8$.}
\label{appendix: observable for 5 models}

\begin{figure}[H]
\centering
\captionsetup{justification=raggedright, singlelinecheck=false, format=plain}

\includegraphics[width=1.\linewidth]{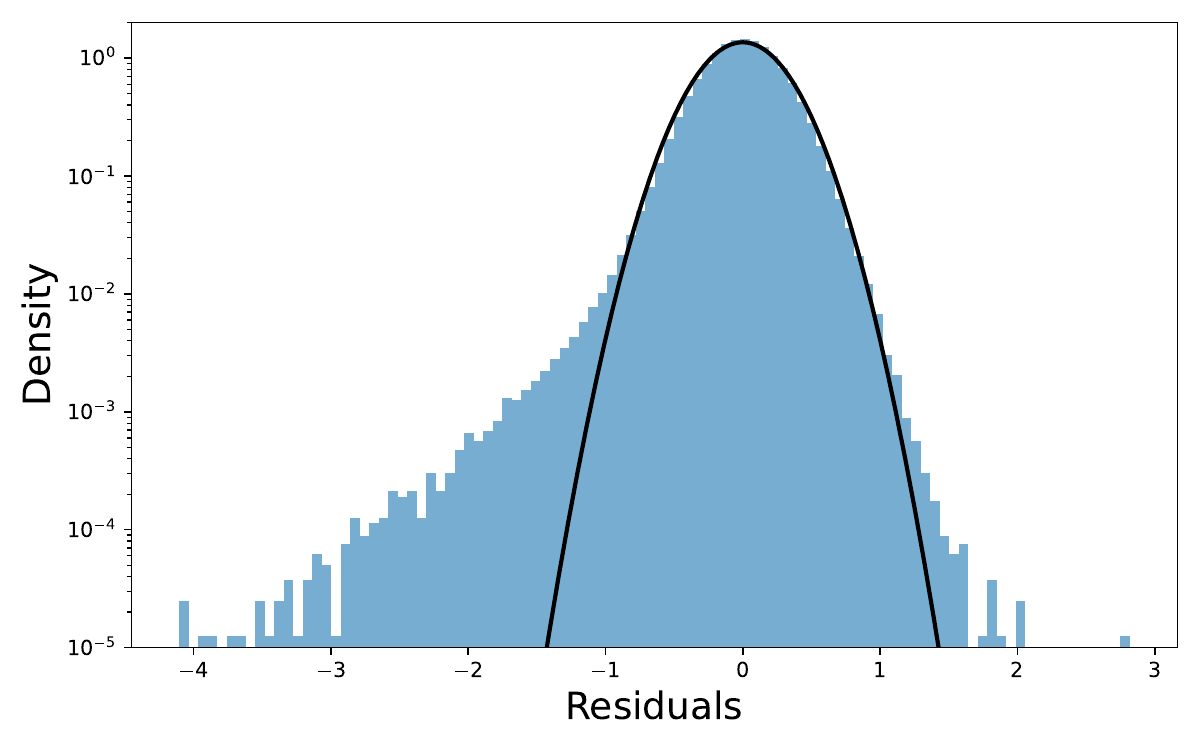}

\includegraphics[width=1.\linewidth]{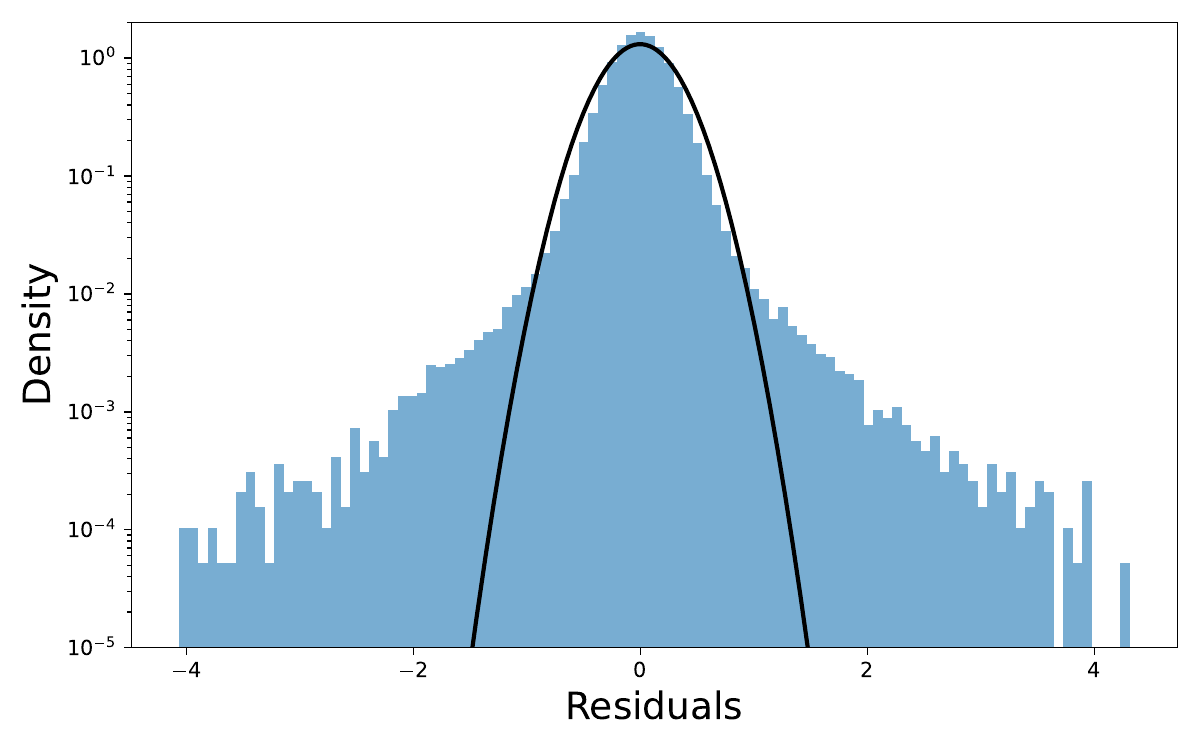}
\caption{
    Density of the residuals lin-log scale PEM $N=5$. 
    Upper panel: Ferromagnetic phase $\beta = 0.7$. Solid black line is the Gaussian fit with $\mu = 0.$ and $\sigma = 0.3$. 
    Lower panel: Paramagnetic phase $\beta =0.5$. Solid black line is the Gaussian fit with $\mu = 0.$ and $\sigma = 0.3.$
    }
\label{residuals_gaussian_5}
\end{figure}

\begin{figure}[h!]
\centering
\captionsetup{justification=raggedright, singlelinecheck=false, format=plain}

\includegraphics[width=1.\linewidth]{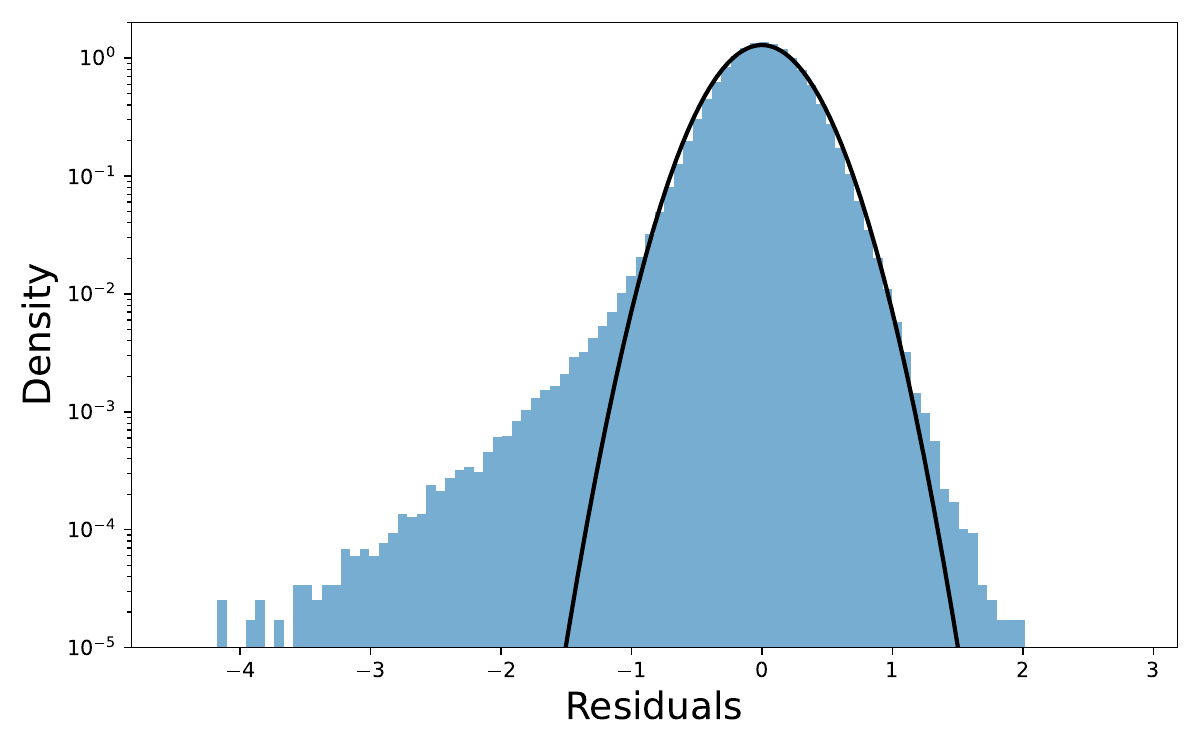}

\includegraphics[width=1.\linewidth]{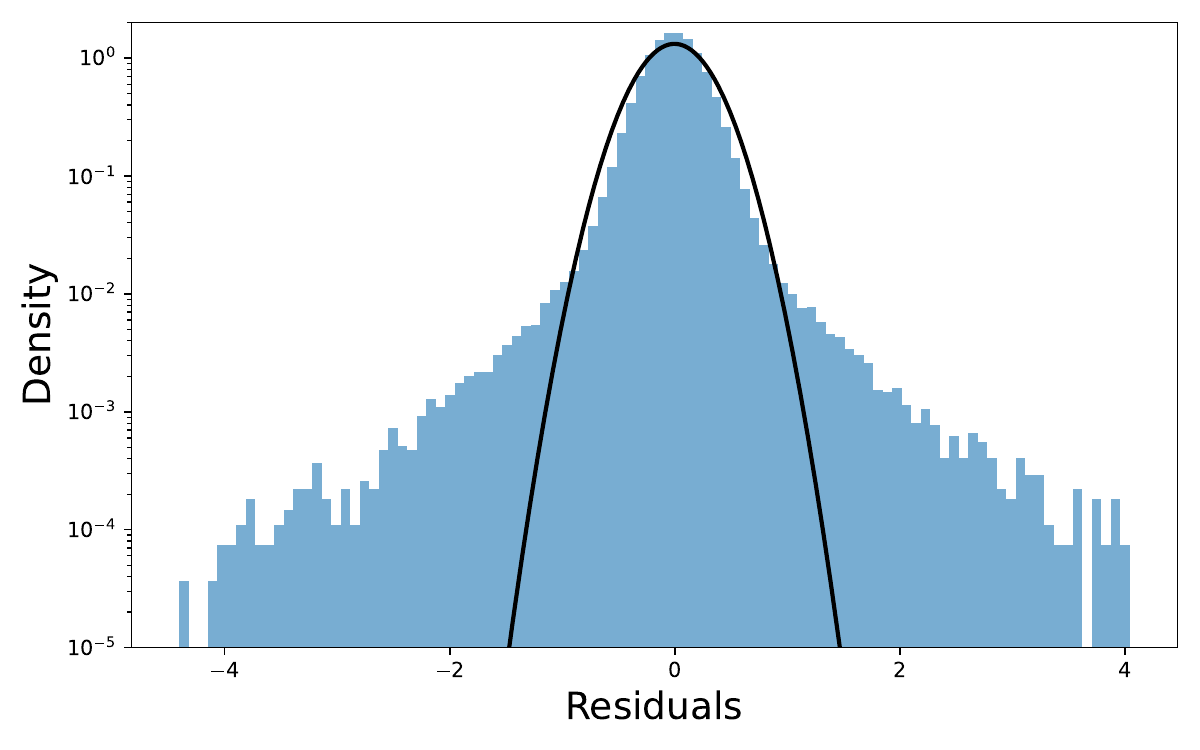}
\caption{
    Density of the residuals lin-log scale PEM $N=7$.
    Upper panel: Ferromagnetic phase $\beta = 0.7$. Solid black line is the Gaussian fit with $\mu = 0.$ and $\sigma = 0.3$. 
    Lower panel: Paramagnetic phase $\beta =0.5$. Solid black line is the Gaussian fit with $\mu = 0.$ and $\sigma = 0.3.$
    }
\label{residuals_gaussian_7}
\end{figure}

\begin{figure}[h!]
\centering
\captionsetup{justification=raggedright, singlelinecheck=false, format=plain}

\includegraphics[width=1.\linewidth]{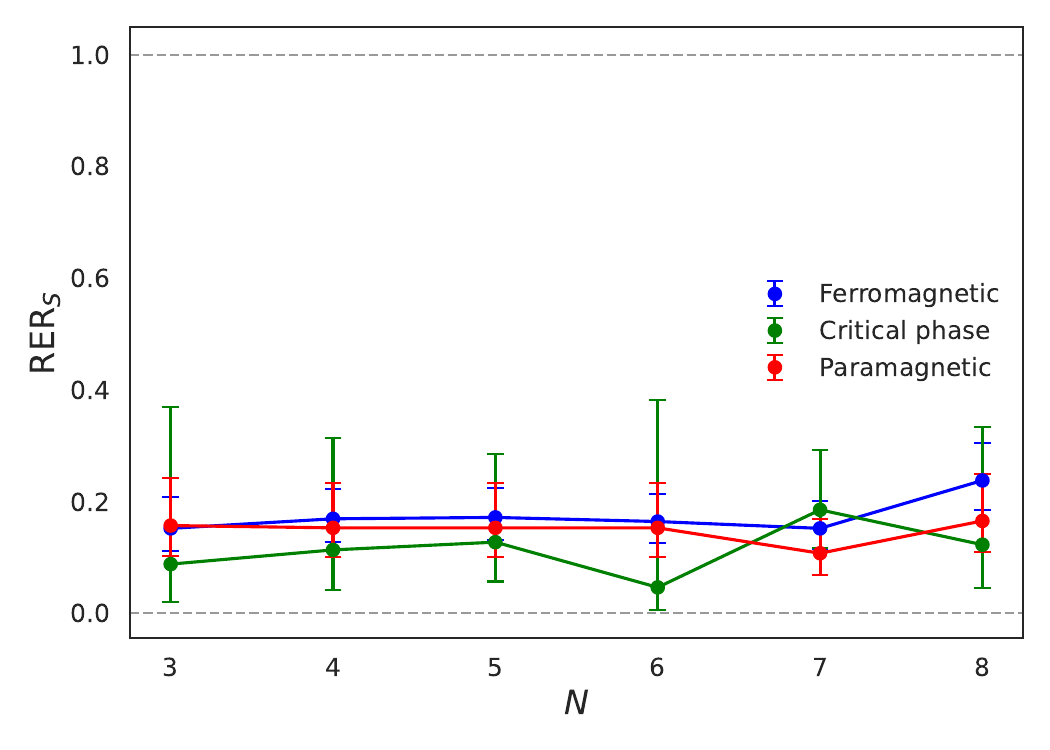}

\includegraphics[width=1.\linewidth]{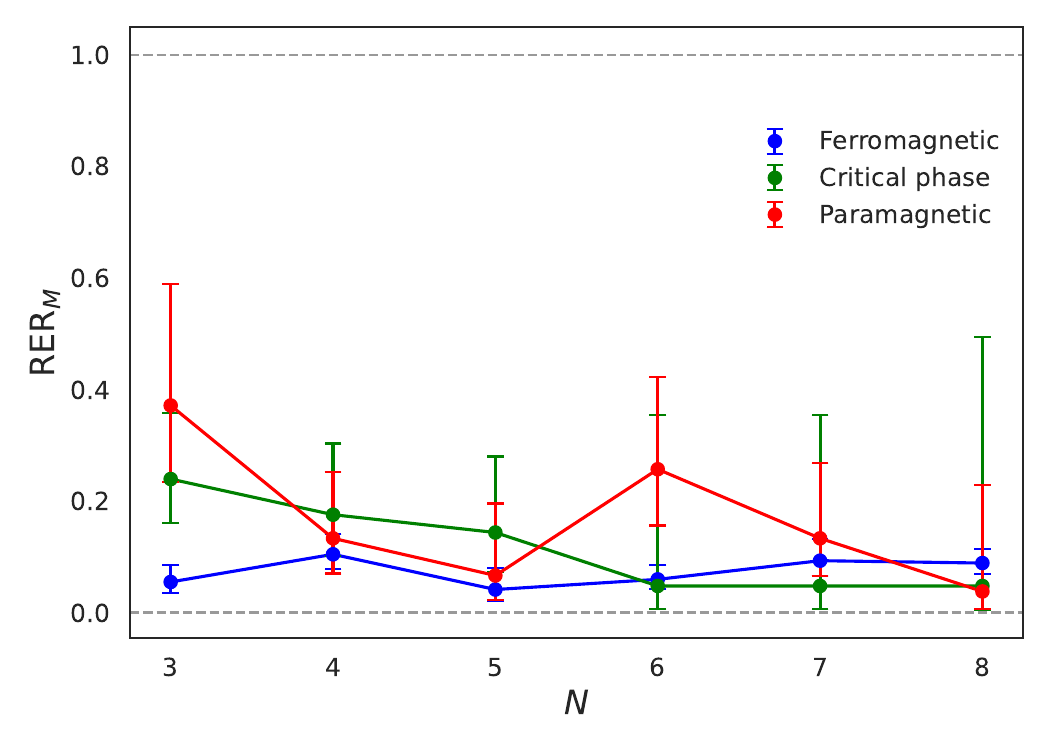}
\caption{ 
        Relative error ratios for action and magnetization across phases. For each ensemble size $N$, $1000$ independent samples were collected in the ferromagnetic and paramagnetic phases, and 500 in the critical phase, using $20$ independent Markov chains. Measurements were recorded every $46080$ steps in the ferromagnetic and paramagnetic regimes, and every $92160$ steps near criticality. The error bars denote one standard deviation around the mean across independent chains, propagated through the relative error ratio.}

\label{relative_error_ratio}
\end{figure}

In this section, we first present the residuals for $N=5$ and $N=7$ in Fig.~\ref{residuals_gaussian_5} and Fig.~\ref{residuals_gaussian_7}, where the behavior of the tails remains largely unchanged. Fig.~\ref{relative_error_ratio} then illustrates the relative error ratio ($\mathrm{RER}$), defined as the ratio of the PEM estimation error to the RCNN estimation error relative to the ground truth, for ensemble sizes ranging from $N=3$ to $N=8$. In the case of the computation of the action, the relative error ratio is defined as
\begin{equation}
    \mathrm{RER}_S = \frac{\left|\mathbb{E}_{\rm GT}[S[\phi]] - \mathbb{E}_{\rm PEM}[S[\phi]]\right|}{\left|\mathbb{E}_{\rm GT}[S[\phi]] - \mathbb{E}_{\rm RCNN}[S[\phi]]\right|}
\end{equation}
and for the magnetization it is given by
\begin{equation}
    \mathrm{RER}_M = \frac{\left|\mathbb{E}_{\rm GT}[M[\phi]] - \mathbb{E}_{\rm PEM}[M[\phi]]\right|}{\left|\mathbb{E}_{\rm GT}[M[\phi]] - \mathbb{E}_{\rm RCNN}[M[\phi]]\right|}.
\end{equation}
Here, $\mathbb{E}_{\rm GT}[\cdot]$, $\mathbb{E}_{\rm PEM}[\cdot]$, and $\mathbb{E}_{\rm RCNN}[\cdot]$ denote the sample mean of the observable over the configurations generated by the Markov chain when using the ground truth Metropolis simulation (with the exact action), the PEM, and the single RCNN, respectively.

\begin{figure*}[ht!]
    \centering
    \captionsetup{justification=raggedright, singlelinecheck=false, format=plain}

    \begin{subfigure}[t]{0.48\textwidth}
        \includegraphics[width=\linewidth]{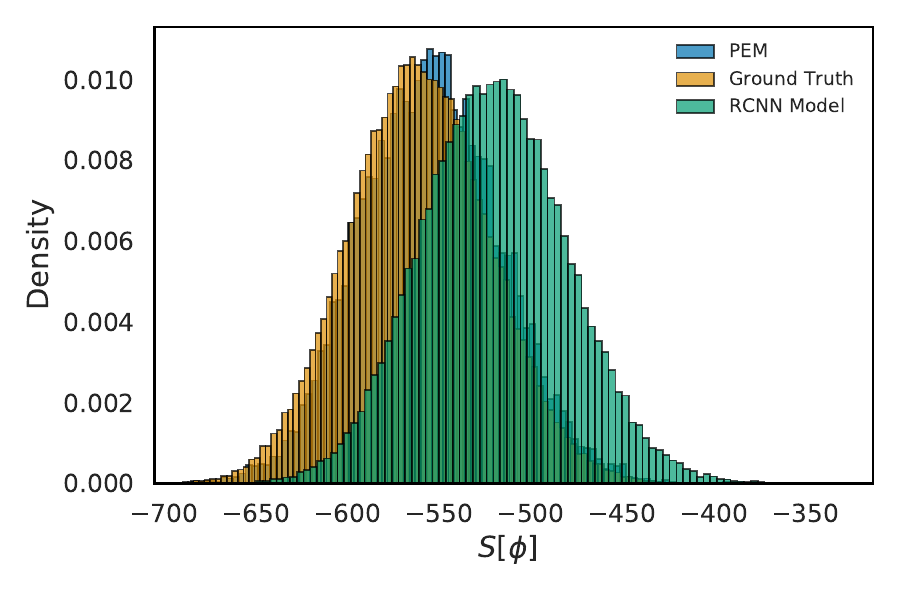}

        \label{fig:mh_ferro}
    \end{subfigure}
    \hfill
    \begin{subfigure}[t]{0.48\textwidth}
        \includegraphics[width=\linewidth]{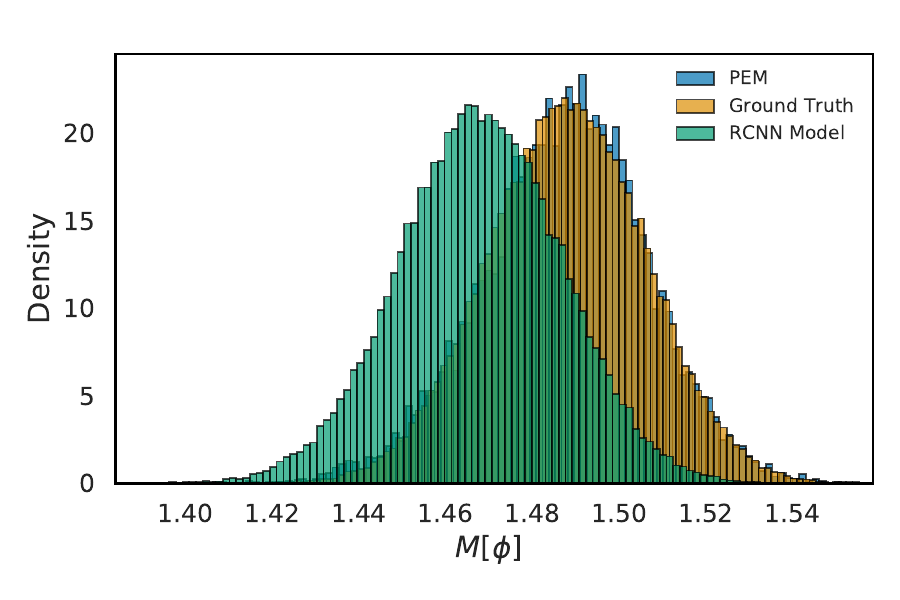}

        \label{fig:mala_ferro}
    \end{subfigure}
    \hfill
    \begin{subfigure}[t]{0.48\textwidth}
        \includegraphics[width=\linewidth]{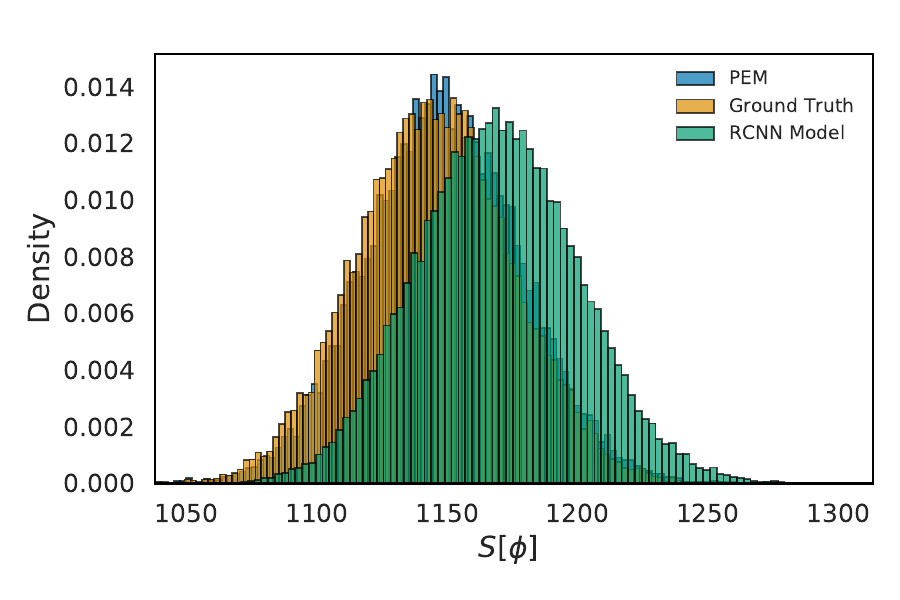}

        \label{fig:mh_param}
    \end{subfigure}
    \hfill
    \begin{subfigure}[t]{0.48\textwidth}
         \includegraphics[width=\linewidth]{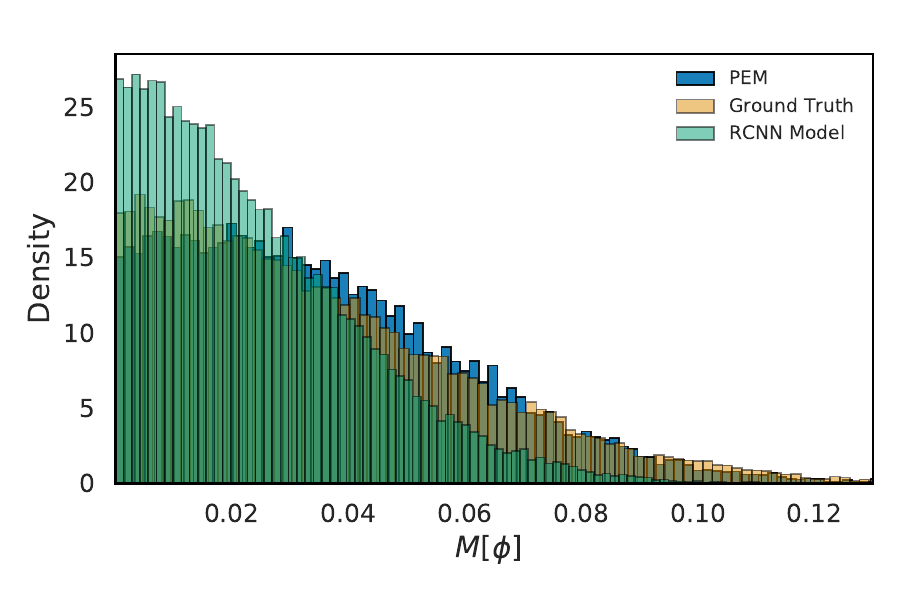}

        \label{fig:mh_algo}
    \end{subfigure}

    \hfill
    \begin{subfigure}[t]{0.48\textwidth}
        \includegraphics[width=\linewidth]{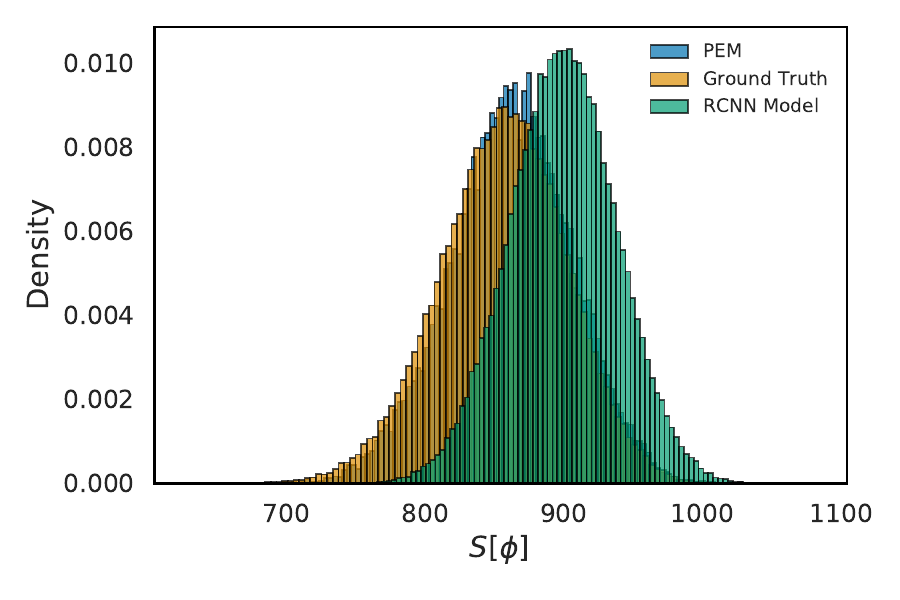}

        \label{fig:mh_algo}
    \end{subfigure}
    \hfill
    \begin{subfigure}[t]{0.48\textwidth}
        \includegraphics[width=\linewidth]{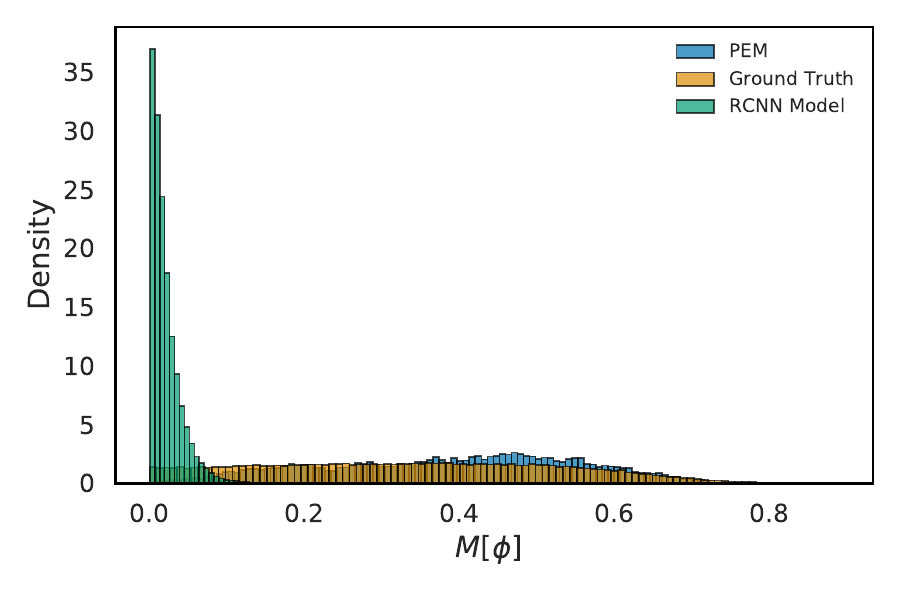}

        \label{fig:mh_algo}
    \end{subfigure}

    \caption{Density of equilibrated, independent configurations of size $48 \times 48$ for the $\phi^4$ model, obtained using RCNN (green) and PEM with $N=5$ (blue). For the Ferromagnetic phase at $\beta=0.7$ and the Paramagnetic phase at $\beta=0.5$, we show histograms based on $1000$ independent configurations (first four panels). Near the phase transition (last two panels), we show histograms based on 500 configurations, collected from 20 independent Markov chains with one sample recorded every $46080$ steps for the Ferromagnetic and Paramagnetic and $92160$ for the phase transition.}
    \label{fig:densities n equal 5}
\end{figure*}

\begin{figure*}[ht!]
    \centering
    \captionsetup{justification=raggedright, singlelinecheck=false, format=plain}

    \begin{subfigure}[t]{0.48\textwidth}
        \includegraphics[width=\linewidth]{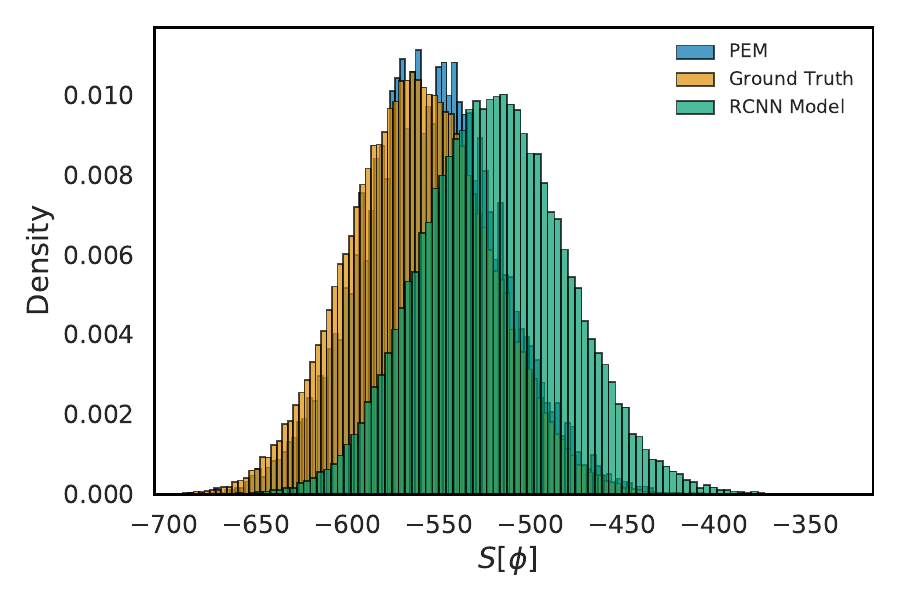}

        \label{fig:mh_ferro}
    \end{subfigure}
    \hfill
    \begin{subfigure}[t]{0.48\textwidth}
        \includegraphics[width=\linewidth]{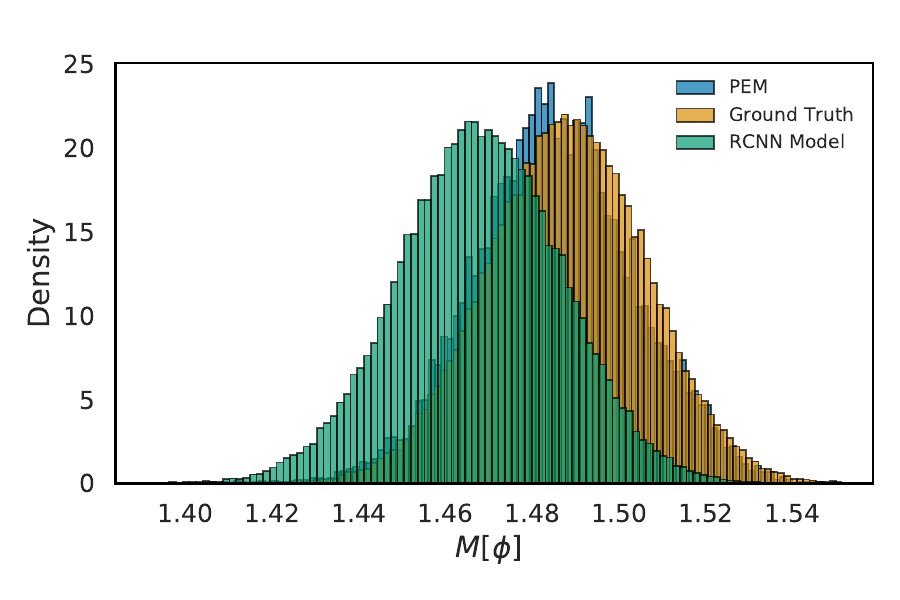}

        \label{fig:mala_ferro}
    \end{subfigure}
    \hfill
    \begin{subfigure}[t]{0.48\textwidth}
        \includegraphics[width=\linewidth]{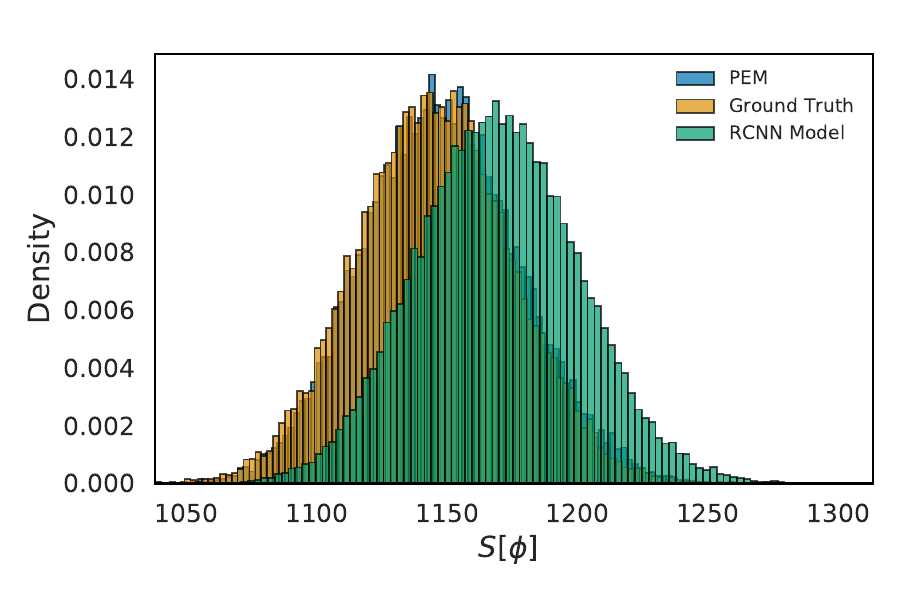}

        \label{fig:mh_param}
    \end{subfigure}
    \hfill
    \begin{subfigure}[t]{0.48\textwidth}
          \includegraphics[width=\linewidth]{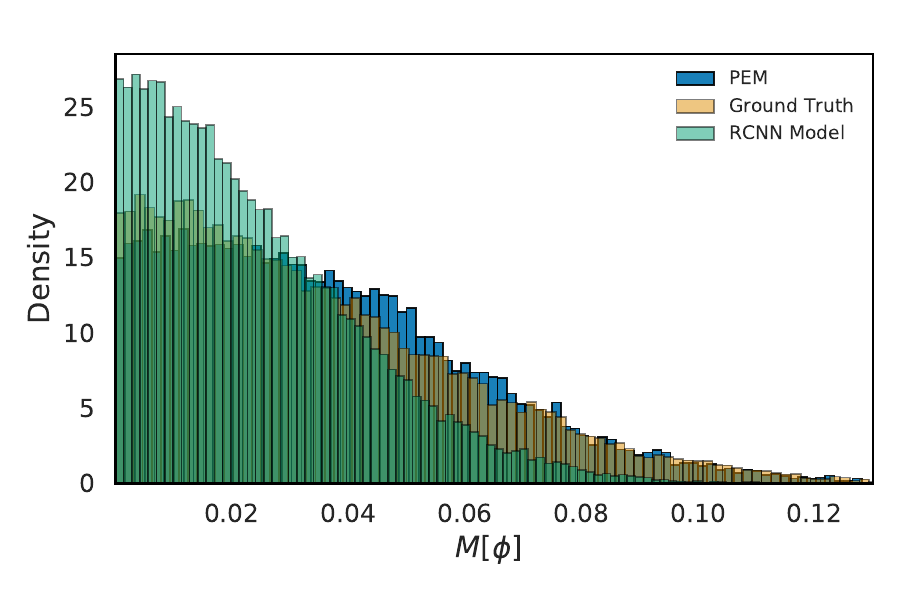}

        \label{fig:mh_algo}
    \end{subfigure}

    \hfill
    \begin{subfigure}[t]{0.48\textwidth}
        \includegraphics[width=\linewidth]{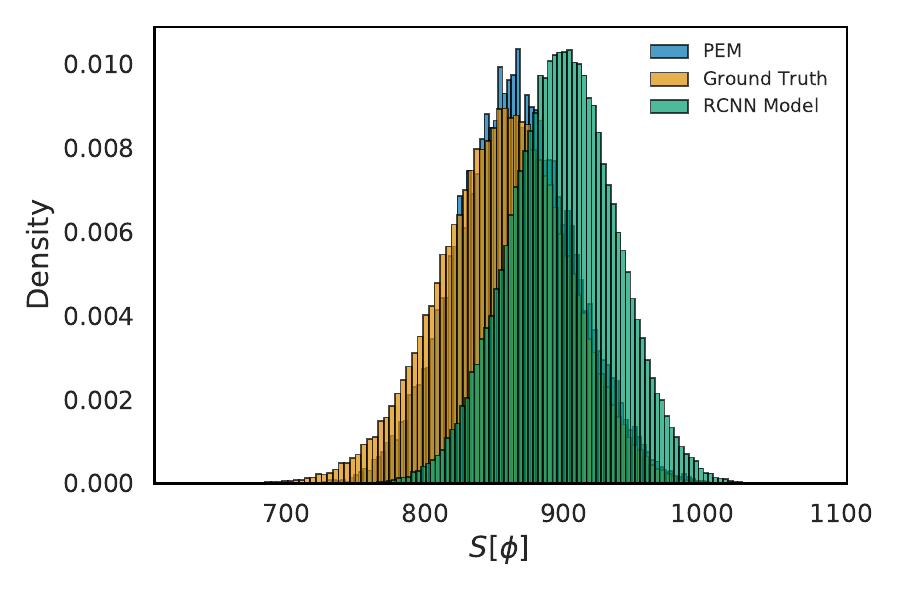}

        \label{fig:mh_algo}
    \end{subfigure}
    \hfill
    \begin{subfigure}[t]{0.48\textwidth}
        \includegraphics[width=\linewidth]{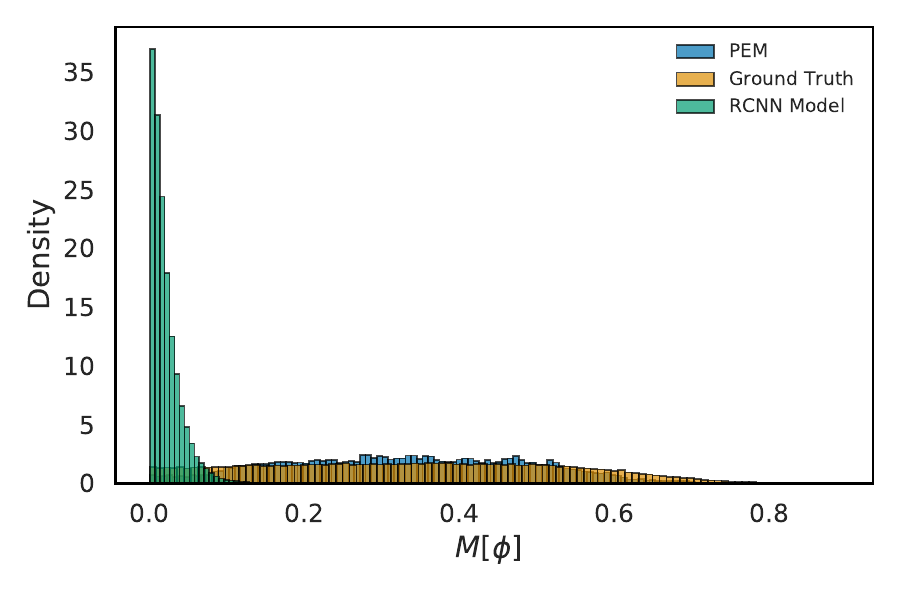}

        \label{fig:mh_algo}
    \end{subfigure}

    \caption{Density of equilibrated, independent configurations of size $48 \times 48$ for the $\phi^4$ model, obtained using RCNN (green) and PEM with $N=7$ (blue). For the Ferromagnetic phase at $\beta=0.7$ and the Paramagnetic phase at $\beta=0.5$, we show histograms based on $1000$ independent configurations (first four panels). Near the phase transition (last two panels), we show histograms based on 500 configurations, collected from 20 independent Markov chains with one sample recorded every 46080 steps for the Ferromagnetic and Paramagnetic and 92160 for the phase transition.}
    \label{fig:densities n equal 7}
\end{figure*}

All measured ratios are consistently smaller than one, confirming that PEM achieves higher accuracy than RCNN for both observables. Moreover, the weak dependence on $N$ shows that the relative error ratio of PEM remains essentially stable as the size of the ensemble increases, indicating that small ensembles are sufficient to realize the method's primary benefits.
Finally, we illustrate the normalized histograms of observables for the PEM with $N=5$ and $N=7$ models in Fig.~\ref{fig:densities n equal 5} and Fig.~\ref{fig:densities n equal 7}, respectively. The results indicate that increasing the ensemble size yields performance comparable to the $N=3$ case.

\captionsetup{margin=0pt}

\captionsetup{margin=0pt}

\begin{figure*}
\centering
\captionsetup{justification=raggedright, singlelinecheck=false}

\begin{subfigure}[t]{0.4\textwidth}
    \centering
    \includegraphics[width=\linewidth]{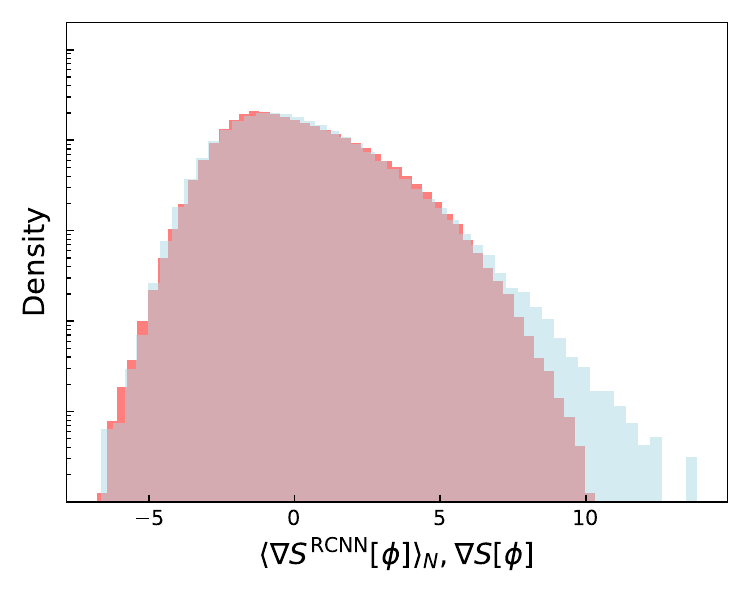}
    \caption{Ferromagnetic phase.}
    \label{fig:grad_ferro}
\end{subfigure}%
\hfill
\begin{subfigure}[t]{0.4\textwidth}
    \centering
    \includegraphics[width=\linewidth]{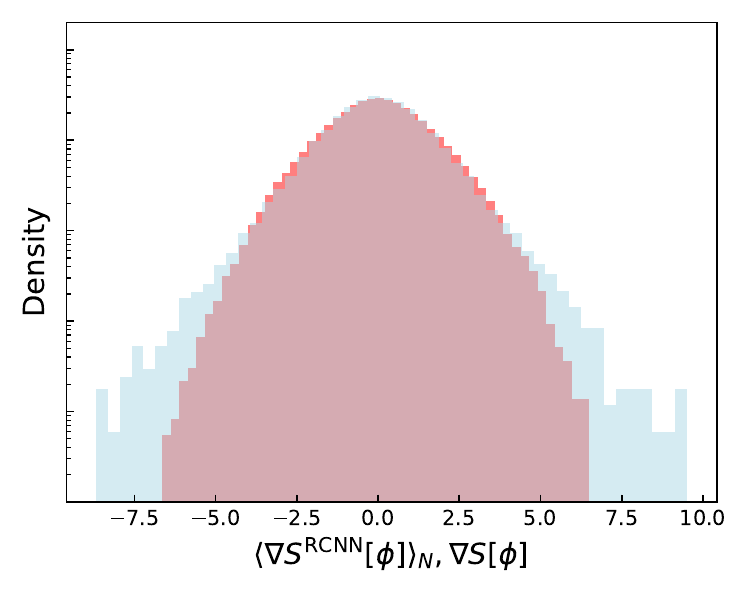}
    \caption{Paramagnetic phase.}
    \label{fig:grad_para}
\end{subfigure}

\caption{Density in lin-log scale of the true gradients versus the ensemble of $N=3$ RCNN-predicted gradients in the ferromagnetic (a) and paramagnetic (b) phases.}
\label{fig:gradients}
\end{figure*}
\setlength{\dbltextfloatsep}{5pt plus 1pt minus 10pt} 

\captionsetup{margin=0pt}
\begin{figure*}[ht!]
    \centering
    \begin{tikzpicture}[
        % GLOBAL SETTINGS
        node distance=0.5cm and 0.3cm, % Vertical and Horizontal spacing
        scale=0.9, transform shape,    % Scaled to fit column
        % STYLES
        layer/.style={rectangle, draw, minimum height=0.7cm, minimum width=1.2cm, font=\sffamily\footnotesize, align=center},
        input/.style={layer, fill=yellow!30},
        conv/.style={layer, fill=orange!30},
        bn/.style={layer, fill=blue!20},
        relu/.style={layer, fill=red!20},
        add/.style={circle, draw, fill=gray!20, inner sep=0pt, minimum size=0.5cm},
        output/.style={layer, fill=green!30},
        skip/.style={layer, fill=gray!30, dashed, font=\scriptsize},
        arrow/.style={->, thick, >=stealth}
    ]

    % --- 1. INPUT ---
    \node[input] (input) {Input $\phi$};

    % --- 2. INITIAL BLOCK ---
    \node[conv, below=0.6cm of input] (init_conv) {Conv\\1$\to$4};
    \node[bn, right=0.3cm of init_conv] (init_bn) {BN};
    \node[relu, right=0.3cm of init_bn] (init_relu) {ReLU};
    
    \draw[arrow] (input) -- (init_conv);
    \draw[arrow] (init_conv) -- (init_bn);
    \draw[arrow] (init_bn) -- (init_relu);

    % --- 3. RES BLOCK 1 (Projection) ---
    % Note: Increased 'below' distance to 1.5cm to give space for previous arrows if needed
    \node[conv, below=1.2cm of init_conv] (rb1_c1) {Conv\\4$\to$8};
    \node[bn, right=of rb1_c1] (rb1_b1) {BN};
    \node[relu, right=of rb1_b1] (rb1_r1) {ReLU};
    \node[conv, right=of rb1_r1] (rb1_c2) {Conv\\8$\to$8};
    \node[bn, right=of rb1_c2] (rb1_b2) {BN};
    \node[add, right=0.3cm of rb1_b2] (rb1_add) {+};
    \node[relu, right=0.3cm of rb1_add] (rb1_out) {ReLU};

    % Connections Main Path
    % Connect from Init ReLU down to start of RB1
    \draw[arrow] (init_relu.south) -- ++(0,-0.6) -| (rb1_c1.north);
    \draw[arrow] (rb1_c1) -- (rb1_b1);
    \draw[arrow] (rb1_b1) -- (rb1_r1);
    \draw[arrow] (rb1_r1) -- (rb1_c2);
    \draw[arrow] (rb1_c2) -- (rb1_b2);
    \draw[arrow] (rb1_b2) -- (rb1_add);
    \draw[arrow] (rb1_add) -- (rb1_out);

    % SKIP CONNECTION (PROJECTION)
    % 1. Move it further LEFT (1.2cm) to avoid overlap
    \node[skip, left=1.2cm of rb1_c1] (rb1_proj) {Proj\\1$\times$1};
    
    % 2. Draw from Init to Projection
    \draw[arrow] (init_relu.south) -- ++(0,-0.6) -| (rb1_proj.north);
    
    % 3. Draw from Projection to Adder (UNDERNEATH)
    % "++(0,-1.0)" moves the line down 1.0cm below the projection before going right
    \draw[arrow] (rb1_proj.south) -- ++(0,-1.0) -| (rb1_add.south);

    % --- 4. RES BLOCK 2 (Identity) ---
    % Increased 'below' to 1.8cm to make room for the skip line above it
    \node[conv, below=1.8cm of rb1_c1] (rb2_c1) {Conv};
    \node[bn, right=of rb2_c1] (rb2_b1) {BN};
    \node[relu, right=of rb2_b1] (rb2_r1) {ReLU};
    \node[conv, right=of rb2_r1] (rb2_c2) {Conv};
    \node[bn, right=of rb2_c2] (rb2_b2) {BN};
    \node[add, right=0.3cm of rb2_b2] (rb2_add) {+};
    \node[relu, right=0.3cm of rb2_add] (rb2_out) {ReLU};

    % Main Path
    \draw[arrow] (rb1_out.south) -- ++(0,-0.6) -| (rb2_c1.north);
    \draw[arrow] (rb2_c1) -- (rb2_b1);
    \draw[arrow] (rb2_b1) -- (rb2_r1);
    \draw[arrow] (rb2_r1) -- (rb2_c2);
    \draw[arrow] (rb2_c2) -- (rb2_b2);
    \draw[arrow] (rb2_b2) -- (rb2_add);
    \draw[arrow] (rb2_add) -- (rb2_out);

    % Skip Connection (Identity)
    % Goes left, down, then under
    \draw[arrow, dashed] ($(rb1_out.south) + (0,-0.6)$) -| ($(rb2_c1.west) - (0.5,0)$) |- (rb2_add.south);

    % --- 5. RES BLOCK 3 (Identity) ---
    % Increased 'below' to 1.8cm
    \node[conv, below=1.8cm of rb2_c1] (rb3_c1) {Conv};
    \node[bn, right=of rb3_c1] (rb3_b1) {BN};
    \node[relu, right=of rb3_b1] (rb3_r1) {ReLU};
    \node[conv, right=of rb3_r1] (rb3_c2) {Conv};
    \node[bn, right=of rb3_c2] (rb3_b2) {BN};
    \node[add, right=0.3cm of rb3_b2] (rb3_add) {+};
    \node[relu, right=0.3cm of rb3_add] (rb3_out) {ReLU};

    % Main Path
    \draw[arrow] (rb2_out.south) -- ++(0,-0.6) -| (rb3_c1.north);
    \draw[arrow] (rb3_c1) -- (rb3_b1);
    \draw[arrow] (rb3_b1) -- (rb3_r1);
    \draw[arrow] (rb3_r1) -- (rb3_c2);
    \draw[arrow] (rb3_c2) -- (rb3_b2);
    \draw[arrow] (rb3_b2) -- (rb3_add);
    \draw[arrow] (rb3_add) -- (rb3_out);

    % Skip Connection
    \draw[arrow, dashed] ($(rb2_out.south) + (0,-0.6)$) -| ($(rb3_c1.west) - (0.5,0)$) |- (rb3_add.south);

    % --- 6. OUTPUT ---
    \node[conv, below=1.5cm of rb3_c1] (out_conv) {Conv\\8$\to$1};
    \node[output, right=0.5cm of out_conv] (output) {Output $\nabla S^{\rm ARCNN}[\phi]$};

    \draw[arrow] (rb3_out.south) -- ++(0,-0.6) -| (out_conv.north);
    \draw[arrow] (out_conv) -- (output);
    
    % Labels
    \node[anchor=east, font=\scriptsize\bfseries, color=gray] at ($(rb1_proj.west) + (-0.2,0)$) {ResBlock 1};
    \node[anchor=east, font=\scriptsize\bfseries, color=gray] at ($(rb2_c1.west) - (0.8,0)$) {ResBlock 2};
    \node[anchor=east, font=\scriptsize\bfseries, color=gray] at ($(rb3_c1.west) - (0.8,0)$) {ResBlock 3};

    \end{tikzpicture}
    \caption{Augmented RCNN architecture. BN denotes Batch Normalization, and the final two residual blocks contain the same number of parameters as the first residual block.}
    \label{fig:res_cnn_final}
\end{figure*}

\begin{figure*}
\centering
\captionsetup{justification=raggedright, singlelinecheck=false}

\begin{subfigure}[t]{0.48\textwidth}
    \centering
    \includegraphics[width=\linewidth]{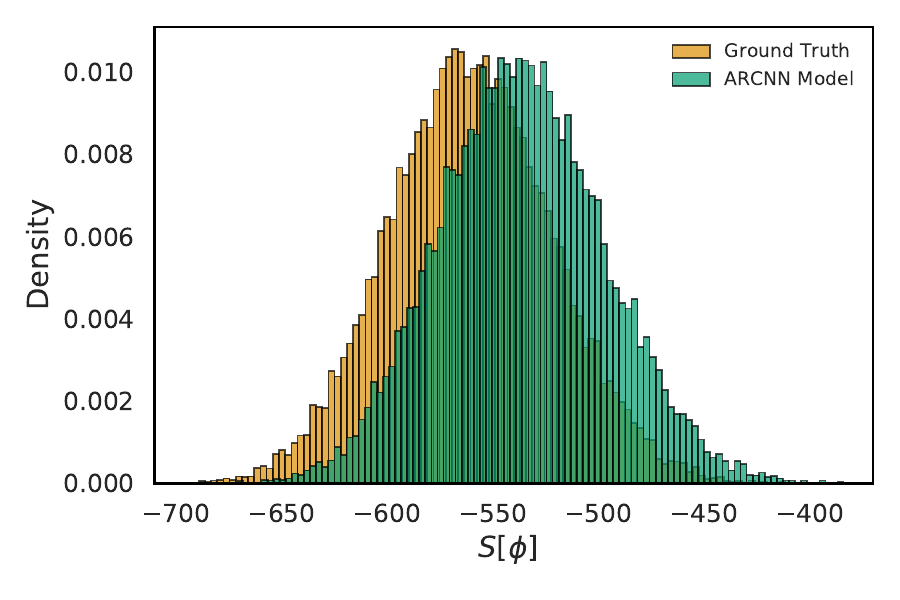}
    \label{fig:arc_actions}
\end{subfigure}%
\hfill
\begin{subfigure}[t]{0.48\textwidth}
    \centering
    \includegraphics[width=\linewidth]{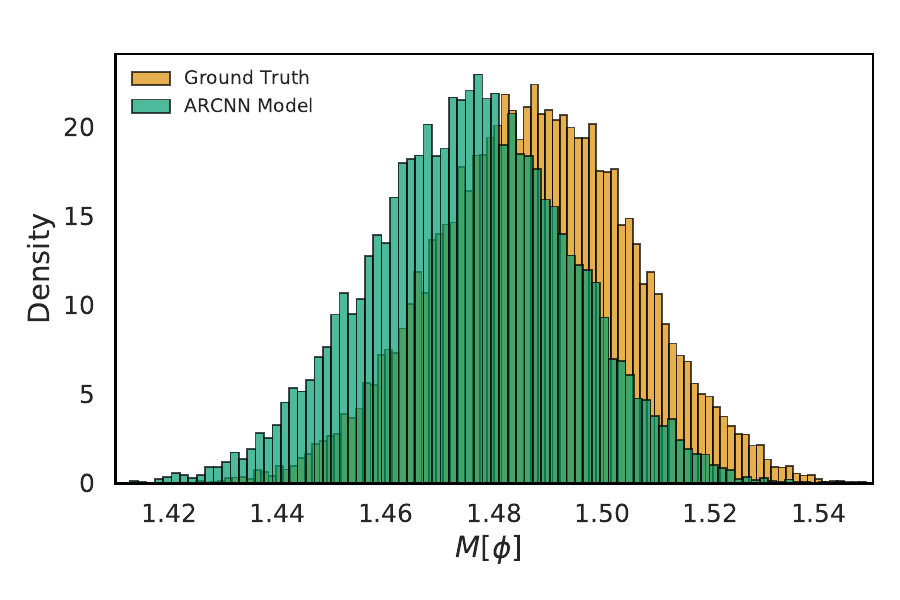}
    \label{fig:arc_mag}
\end{subfigure}

\caption{Densities of $1000$ independent, equilibrated configurations of size $48 \times 48$ in the ferromagnetic phase ($\beta=0.7$) of the $\phi^4$ model. Samples were obtained from 20 independent Markov chains, recording one configuration every $46080$ steps. The ARCNN (green) fails to correctly reproduce the ferromagnetic configurations.}
\label{fig:arcnn_densities}
\end{figure*}

\section{Histograms of the gradients.}
\label{appendix of gradients}
In Fig.~\ref{fig:grad_ferro} and \ref{fig:grad_para}, we compare densities of the true gradient components with those of the ensemble RCNN  $\langle \nabla S^{\rm RCNN}[\phi] \rangle_N$. While  Fig.~\ref{performance_grad} suggests that the RCNN predictions are consistent
with the true gradient behavior, a more detailed comparison through the
distributions shows that the RCNN fails to capture important details about the
gradient.

\section{ Effect of Model Capacity on Sampling Performance.}

\label{appendix of advanced model}

In this appendix, we address the question of whether increasing the capacity and computational cost of a single RCNN model can reproduce the stability and accuracy achieved by the PEM.

To this end, we trained a new RCNN architecture with substantially increased number of trainable parameters, while keeping the total number of training samples fixed and taken from the ferromagnetic phase, in order to ensure a fair comparison with the ensemble approach. In particular, we considered the augmented RCNN architecture illustrated in Fig.~\ref{fig:res_cnn_final}. We refer to this model as ARCNN, where “A” denotes augmented.

In Table~\ref{table:model_comparison}, we report the number of trainable parameters (\textbf{Params}), along with the training time (\textbf{Train}) and the average inference time (\textbf{Infer.}), computed over 100 independent runs, together with the mean and standard deviation. All experiments were performed using an NVIDIA Tesla T4 GPU. The ARCNN was trained on the same total dataset as the ensemble used in the PEM and required 50 additional training epochs to reach optimal performance.

\begin{table}[H]
    \centering
    \renewcommand{\arraystretch}{1.6}
    \caption{Model Comparison}
    \begin{tabular}{c|ccc}
        \hline
        & \textbf{RCNN} & \textbf{PEM N=3} & \textbf{ARCNN} \\
        \hline
        \textbf{Params} 
        & 473 
        & 1419
        & 3473 \\
        
        \textbf{Train (s)} 
        & 2.2  
        & 6.5  
        & 6.9  \\
        
        \textbf{Infer. (ms)} 
        & $0.6 \pm 0.1 $ 
        & $1.7 \pm 0.2 $
        & $1.5 \pm 0.2$ \\
        \hline
    \end{tabular}
    \label{table:model_comparison}
\end{table}

The sampling performance of this larger model was evaluated using the same Monte Carlo procedure as in the main text. Fig.~\ref{fig:arcnn_densities} shows the resulting
distributions of the action $S[\phi]$ and the magnetization $M[\phi]$. Despite the significantly increased model capacity and computational cost, the augmented architecture does not exhibit an improvement over the baseline RCNN. On the contrary, the model displays clear deviations from the ground truth distributions.

These results demonstrate that simply increasing the number of trainable parameters and layers in a single RCNN is insufficient to recover the robustness and accuracy of the ensemble-based PEM. This highlights that the improved performance of the PEM originates from the reduction of gradient noise through model averaging, rather than from an increased model capacity alone.

\end{document}